\documentclass[epj]{svjour}
\usepackage{graphicx,amsmath,amssymb}

\usepackage[unicode=true,
 bookmarks=true,bookmarksnumbered=false,bookmarksopen=false,
 breaklinks=false,pdfborder={0 0 1},backref=false,colorlinks=false]
 {hyperref}
\hypersetup{pdftitle={Equilibrium States of the Random Field XY},
 pdfauthor={D. A. Garanin, E. M. Chudnovsky, and T. Proctor},
 pdfkeywords={random field XY model}}
\usepackage{breakurl}

\newcommand{\lyxdot}{.}

\begin{document}

\title{Ordered vs Disordered States of the Random-Field Model in Three Dimensions}

\author{D. A. Garanin and E. M. Chudnovsky}

\institute{Physics Department, Lehman College, City University of New York \\
 250 Bedford Park Boulevard West, Bronx, New York 10468-1589, USA}

\date{\today}
\abstract{
We report numerical investigation of the glassy behavior of random-field
exchange models in three dimensions. Correlation of energy with the
magnetization for different numbers of spin components has been studied.
There is a profound difference between the models with two and three
spin components with respect to the stability of the magnetized state due to the different kinds of singularities:
vortex loops and hedgehogs, respectively. Memory
effects pertinent to such states have been investigated. Insight into
the mechanism of the large-scale disordering is provided by numerically
implementing the Imry-Ma argument in which the spins follow the random
field averaged over correlated volumes. Thermal stability of the magnetized
states is investigated by the Monte Carlo method.
} 

\maketitle

\section{Introduction}

\label{introduction}

Recently, it has been argued that the behavior of the order parameter
in the random field (RF) model described by the Hamiltonian
\begin{equation}
{\cal H}=\int d^{d}r\left[\frac{\alpha_{e}}{2}(\partial_{\mu}{\bf S})\cdot(\partial_{\mu}{\bf S})-{\bf h}\cdot{\bf S}\right]\label{H-continuous}
\end{equation}
is controlled by topology. \cite{PGC-PRL2014} Here $\alpha_{e}$
is the exchange stiffness and ${\bf S}$ is the $n$-component fixed-length
vector field (e.g., spin density of constant length $S_{0}$) interacting
with the $n$-component RF ${\bf h}({\bf r})$ in $d$ dimensions.
Summation over repeated indices is assumed. At $n>d+1$ the behavior
of the system is fully reversible with the non-ferromagnetic ground
state and exponential decay of spin-spin correlations, while at $n\leq d+1$
the system exhibits glassy behavior with its state determined by the
initial condition. \cite{GCP-EPL2013} In the latter case, the fully
ordered initial state relaxes to the partially disordered state possessing a non-zero magnetization
that we call the F-state. For the $xy$ model ($n=2$) in three dimensions
the properties of that state have been studied in Ref.\ \cite{GCP-PRB13}.
In this paper we focus on the comparative study of the F-state in
a $3d$ $xy$ model and in a $3d$ Heisenberg model ($n=3$) with
the goal to shed more light on the glassy properties of the RF system.

The problem has a long history. More than forty years ago Larkin,
within a conceptually similar model, argued that randomly positioned
pinning centers destroy the translational order in a flux-line lattice.
\cite{Larkin-JETP1970} Correlations associated with the translational
order in flux lattices are of practical interest because they define
the size of the vortex bundle that gets depinned by the transport
current, which, in turn, determines the critical current. \cite{Blatter-RMP1994}
Related to this effect is a more general qualitative argument suggested
by Imry and Ma. \cite{Imry-Ma-PRL1975} It states that a static RF,
regardless of strength, destroys the long-range order associated with
a continuous-symmetry order parameter below $d=4$ spatial dimensions.
Aizenman and Wehr \cite{Aizenman-Wehr-PRL,Aizenman-Wehr-CMP} provided
a mathematical argument that is considered to be a rigorous proof
of the Imry-Ma statement. According to this statement the directions
of ${\bf S}$ are correlated only within randomly oriented domains
of average size $R_{f}\propto(1/h)^{2/(4-d)}$. Same $R_{f}$ comes
from the Green-function method. \cite{EC-Serota-PRB82} These ideas
have been applied to random magnets, \cite{Pelcovits,Patterson,Aharony,EC-PRB86}
disordered antiferromagnets, \cite{Fishman} spin-glasses, \cite{Binder}
arrays of magnetic bubbles, \cite{bubbles} superconductors, \cite{EC-PRB1991,Blatter-RMP1994}
charge-density waves, \cite{Efetov-77} liquid crystals, \cite{LC}
superconductor-insulator transition, \cite{SI} and superfluid $^{3}$He-A
in aerogels. \cite{Volovik,Li}

In early 1980s the renormalization group treatments of the problem
by Cardy and Ostlund \cite{Cardy-PRB1982} and by Villain and Fernandez
\cite{Villain-ZPB1984} questioned the validity of Larkin-Imry-Ma
(LIM) argument for distances $R\gtrsim R_{f}$. The application of
scaling and replica-symmetry breaking arguments to statistical mechanics
of flux lattices, \cite{Nattermann,Kierfield,Korshunov-PRB1993,Giamarchi-94,Giamarchi-95,Nattermann-2000,Feldman,LeDoussal-Wiese-PRL2006,LeDoussal-PRL2006,LeDoussal-PRL07,Bogner}
as well as the variational approach, \cite{Orland-EPL,Garel-PRB}
yielded the power-law decay of correlations at large distances. These
findings suggested that ordering could be more robust against weak
static randomness than expected from the LIM theory. Such a quasiordered
phase, presumed to be vortex-free in spin systems and dislocation-free
in flux lattices, received the name of a Bragg glass. Fisher \cite{Fisher-PRL1997}
called it an ``elastic glass'' and argued that the energy associated
with vortex loops prevents the $xy$ $3d$ RF system from complete
disordering.

Numerical evidence of the Bragg/elastic glass has been inconclusive
so far. Early numerical work on $1d$ (Ref. \cite{DC-PRB1991})
and $2d$ (Ref. \cite{Dieny-PRB1990}) spin systems with quenched
random anisotropy established strong non-equilibrium effects, such as magnetic
hysteresis and dependence on the initial conditions. Defect-free spin
models with relatively large RF and random anisotropy have been studied
numerically on small lattices by Fisch. \cite{Fisch} In three dimensions
strong non-equilbrium effects have been reported in Refs.\ \cite{Gingras-Huse-PRB1996,GCP-EPL2013,GCP-PRB13}.
Numerical studies of the $2d$ elastic media with pinning centers (that is similar to $2d$ $xy$ model) \cite{Fisher-PRL99} revealed
that pinning creates dislocations and thus destroys the Bragg glass. In
line with the conjecture made in Ref.\ \cite{Fisher-PRL1997},
it was suggested in Ref.\ \cite{GCP-PRB13} that the high
energy cost of vortex loops was preventing the spins in the $xy$
$3d$ RF model from relaxing to a disordered state from the initially
ordered state. At elevated temperatures, however, the numerical evidence
of the power-law decay of correlations in a $2d$ random-field $xy$
model has been recently obtained by Perret et al. \cite{Perret-PRL2012}
In the absence of topological defects, the evidence of the logarithmic
growth of misalignment with the size of the system has been also found
in $2d$ Monte Carlo studies of a crystal layer on a disordered substrate
and for pinned flux lattices. \cite{Zeng,Rieger} The power-law decay
of spin-spin correlations has been reported in Monte Carlo studies
of the RF Heisenberg model, \cite{Itakura-03} as well as for the
$xy$ model. \cite{Itakura-05} As to the real experiments, large
areas of defect-free flux lattices have been reported in Ref.\ \cite{Klein-Nature2001}.
The comparison of such experiments with theory is hampered, however,
by the fact that for a weak disorder the correlation length in $3d$
can be very large, making it difficult to distinguish large defect-free
slightly disordered domains from the Bragg glass.

The paper has the following structure. Some rigorous analytical results
that serve as the test for numerics are given in Section \ref{analytical}.
Numerical method and the model used in computations are specified in Section \ref{sec:numerical}.
Energy of the F-state as a function of the magnetization for different
number of spin components is obtained in Section \ref{E vs m}. Evolution
of the F-state generated by a hysteresis cycle is studied in Section
\ref{hysteresis}. The dependence of the magnetization of the F-state
as function of the strength of the RF is reported in Section \ref{m vs h}.
Numerical implementation of the Imry-Ma argument is developed in Section
\ref{IM}. Microscopic structure of the F-state is discussed in Section
\ref{structure}. Memory of the initial condition that results in
the rotational elasticity of the F-state is demonstrated in Section
\ref{rotational}. Section \ref{temperature} deals with the efect
of finite temperature on the F-state. Our conclusions are given in
Section \ref{conclusions}.

\section{Analytical Results}

\label{analytical}

The discrete counterpart of the Hamiltonian (\ref{H-continuous})
that takes into account the Zeeman interaction of spins with the external
field ${\bf H}$ is given by
\begin{equation}
{\cal H}=-\frac{1}{2}\sum_{ij}J_{ij}{\bf s}_{i}\cdot{\bf s}_{j}-\sum_{i}{\bf h}_{i}\cdot{\bf s}_{i}-{\bf H}\cdot\sum_{i}{\bf s}_{i},\label{eq:ham-discrete}
\end{equation}
where ${\bf s}_{i}$ is a $n$-component constant-length ($\left|{\bf s}_{i}\right|=s$)
spin at the site $i$ of a cubic lattice and ${\bf h}_{i}$ is a quenched
RF at that site. The summation is over the nearest neighbors. The
factor 1/2 in the first term is compensating for the double counting
of the exchange bonds. In what follows we assume nearest-neighbor
exchange. The connection between the parameters of the continuous
and discrete models is $\alpha_{e}=Ja^{d+2}$ and $S_{0}=s/a^{d}$,
with $a$ being the lattice spacing. Eq. (\ref{eq:ham-discrete})
contains the exchange energy per spin of the collinear state, $E_{0}=-3Js^{2}$,
that is not included in Eq. (\ref{H-continuous}).

In this paper we present numerical results on the energy minimization
in Eq. (\ref{eq:ham-discrete}) for the uncorrelated RF,
\begin{equation}
\left\langle h_{i\alpha}h_{j\beta}\right\rangle =\frac{h^{2}}{n}\delta_{\alpha\beta}\delta_{ij},\label{eq:h-corr-ij}
\end{equation}
(Greek indices being the Cartesian components of the vectors) although
computations for a correlated RF have been performed as well. Our
main choice for the numerical work has been a fixed-length RF: $\left|\mathbf{h}_{i}\right|=h=\mathrm{const}$.
No difference has been found for models with a distributed RF strength,
e.g. Gaussian.

Before doing numerical work it is useful to obtain some exact analytical
formulas that can provide the ultimate test for our numerical results.
One such formula describes the short-range behavior of spin-spin correlations.
Choosing in $3d$
\begin{equation}
{\cal H}_{\lambda}={\cal H}-\int d^{3}r\lambda({\bf r}){\bf S}^{2}
\end{equation}
to account for ${\bf S}^{2}=S_{0}^{2}={\rm const}$ by the term containing
a Lagrange multiplier $\lambda({\bf r})$, one obtains the following
extremal equation for ${\bf S}$:
\begin{equation}
\alpha_{e}\nabla^{2}{\bf S}+{\bf h}+2\lambda{\bf S}=0.\label{extremal}
\end{equation}
Multiplying this by ${\bf S}$ we have
\begin{eqnarray}
 &  & \lambda=-\frac{1}{2S^{2}}(\alpha_{e}{\bf S}\cdot\nabla^{2}{\bf S}+{\bf S}\cdot{\bf h})\\
 &  & \alpha_{e}\nabla^{2}{\bf S}-\frac{\alpha_{e}}{S_{0}^{2}}{\bf S}({\bf S}\cdot\nabla^{2}{\bf S})+{\bf h}-\frac{1}{S_{0}^{2}}{\bf S}({\bf S}\cdot{\bf h})=0.\label{equation-S}
\end{eqnarray}

In the second term of Eq.\ (\ref{equation-S})
\begin{equation}
{\bf S}\cdot\nabla^{2}{\bf S}=\partial_{\mu}({\bf S}\cdot\partial_{\mu}{\bf S})-\partial_{\mu}{\bf S}\cdot\partial_{\mu}{\bf S}=-\partial_{\mu}{\bf S}\cdot\partial_{\mu}{\bf S}
\end{equation}
because
\begin{equation}
{\bf S}\cdot\partial_{\mu}{\bf S}=\frac{1}{2}\partial_{\mu}{\bf S}^{2}=0.
\end{equation}
As long as small volumes are concerned, this term is quadratic on
the perturbation of ${\bf S}$ caused by the weak RF and, therefore,
it is small compared to other terms in Eq.\ (\ref{equation-S}) that
are linear on ${\bf h}$. Consequently, at small distances this term
can be safely dropped. Implicit solution of the remaining equation
is
\begin{equation}
{\bf S}({\bf r})=-\frac{1}{\alpha_{e}}\int d^{3}r'G({\bf r}-{\bf r}')\left\{ {\bf h}({\bf r}')-\frac{{\bf S}({\bf r}')[{\bf S}({\bf r}')\cdot{\bf h}({\bf r}')]}{S_{0}^{2}}\right\}
\end{equation}
with $G({\bf r})=-1/(4\pi|{\bf r}|)$ being the Green function of
the $3d$ Laplace equation. Then
\begin{eqnarray}
 &  & \frac{1}{2S_{0}^{2}}\langle[{\bf S}({\bf r}_{1})-{\bf S}({\bf r}_{2})]^{2}\rangle=\nonumber \\
 &  & =\frac{1}{2\alpha_{e}^{2}S_{0}^{2}}\int d^{3}r'\int d^{3}r''[G({\bf r}_{1}-{\bf r}')-G({\bf r}_{2}-{\bf r}')]\times\nonumber \\
 &  & [G({\bf r}_{1}-{\bf r}'')-G({\bf r}_{2}-{\bf r}'')]\langle{\bf g}({\bf r}')\cdot{\bf g}({\bf r}'')\rangle,
\end{eqnarray}
where ${\bf g}\equiv{\bf h}-{\bf S}({\bf S}\cdot{\bf h})/S_{0}^{2}$.

Neglecting the weak correlation between ${\bf h}$ and ${\bf S}$,
with the help of the continuous equivalent of Eq.\ (\ref{eq:h-corr-ij}),
\begin{equation}
\langle h_{\alpha}({\bf r}')h_{\beta}({\bf r}'')\rangle=\frac{h^{2}}{n}\delta_{\alpha\beta}a^{3}\delta({\bf r}'-{\bf r}''),\label{eq:h-corr_continuous}
\end{equation}
one obtains
\begin{equation}
\langle g_{\alpha}({\bf r}')g_{\beta}({\bf r}'')\rangle=\frac{h^{2}}{n}\left(\delta_{\alpha\beta}-\frac{\langle S_{\alpha}S_{\beta}\rangle}{S_{0}^{2}}\right)a^{3}\delta({\bf r}'-{\bf r}'').
\end{equation}
Consequently
\begin{equation}
\langle{\bf g}({\bf r}')\cdot{\bf g}({\bf r}'')\rangle=\frac{h^{2}}{n}(n-1)a^{3}\delta({\bf r}'-{\bf r}'').
\end{equation}
This gives
\begin{eqnarray}
 &  & \frac{1}{2S_{0}^{2}}\langle[{\bf S}({\bf r}_{1})-{\bf S}({\bf r}_{2})]^{2}\rangle=\nonumber \\
 &  & =\frac{h^{2}a^{3}}{2\alpha_{e}^{2}S_{0}^{2}}\left(1-\frac{1}{n}\right)\int d^{3}r[G({\bf r}_{1}-{\bf r})-G({\bf r}_{2}-{\bf r})]^{2}\nonumber \\
 &  & =\frac{h^{2}a^{3}}{8\pi\alpha_{e}^{2}S_{0}^{2}}\left(1-\frac{1}{n}\right)|{\bf r}_{1}-{\bf r}_{2}|=\frac{|{\bf r}_{1}-{\bf r}_{2}|}{R_{f}}
\end{eqnarray}
with \cite{PGC-PRL2014}
\begin{equation}
\frac{R_{f}}{a}=\frac{8\pi\alpha_{e}^{2}S_{0}^{2}}{h^{2}a^{4}(1-1/n)}=\frac{8\pi}{(1-1/n)}\left(\frac{Js}{h}\right)^{2}.\label{Rf-n}
\end{equation}
The weakness of the RF should be measured against the exchange field
$6Js$ created in a $3d$ cubic lattice by the nearest neighbors of
each spin. This can be seen by presenting Eq.\ (\ref{Rf-n}) in the
form
\begin{equation}
\frac{R_{f}}{a}=\frac{2\pi}{9(1-1/n)}\left(\frac{6Js}{h}\right)^{2},\label{Rf-n-6J}
\end{equation}
with the numerical factor in front of $(6Js/h)^{2}$ of order unity,
e.g., $2\pi/6$ for $n=3$.

Noticing that
\begin{equation}
\frac{1}{2S_{0}^{2}}\langle[{\bf S}({\bf r}_{1})-{\bf S}({\bf r}_{2})]^{2}\rangle=1-\frac{1}{S_{0}^{2}}\langle{\bf S}({\bf r}_{1})\cdot{\bf S}({\bf r}_{2})\rangle\label{eq:S-difference_to_CF}
\end{equation}
we finally obtain
\begin{equation}
\langle{\bf s}_{i}\cdot{\bf s}_{j}\rangle=s^{2}\left(1-\frac{|{\bf r}_{i}-{\bf r}_{j}|}{R_{f}}\right).\label{CF-SR}
\end{equation}
This formula is in agreement with the famous Larkin's result, \cite{Larkin-JETP1970}
with Eq.\ (\ref{Rf-n}) providing the correlation length for arbitrary
$n$.

Long-range correlations are difficult to obtain by the above Green-function
method because of the high non-linearity of Eq.\ (\ref{equation-S}).
However, the case of $n=\infty$ permits an exact analytical solution
at all distances due to its equivalence \cite{Stanley} to the mean-spherical
model in which only the volume average of ${\bf S}^{2}$ rather than
the local ${\bf S}^{2}$ is a constant, $V^{-1}\int d^{3}r{\bf S}^{2}=S_{0}^{2}$.
In this case $\lambda$ in Eq.\ (\ref{extremal}) is a constant that
we will write as $\lambda=-\alpha_{e}k^{2}/2$. Then the extremal
equation for ${\bf S}$ is linear
\begin{equation}
(\nabla^{2}-k^{2}){\bf S}=-\frac{1}{\alpha_{e}}{\bf h}\label{linear}
\end{equation}
with a general solution
\begin{equation}
{\bf S}({\bf r})=-\frac{1}{\alpha_{e}}\int d^{3}r'G_{k}({\bf r}-{\bf r}'){\bf h}({\bf r}'),\label{eq:S_via_h_mean_spherical}
\end{equation}
$G_{k}({\bf r})=-e^{-k|{\bf r}|}/(4\pi|{\bf r}|)$ and $G_{k}({\bf q})=-{1}/({q^{2}+k^{2}})$
being the Green function of Eq.\ (\ref{linear}) and its Fourier
transform, respectively. Consequently,
\begin{eqnarray}
 &  & \langle{\bf S}({\bf r}_{1})\cdot{\bf S}({\bf r}_{2})\rangle=\frac{1}{\alpha_{e}^{2}}\int d^{3}rd^{3}r'G_{k}({\bf r}_{1}-{\bf r}')G_{k}({\bf r}_{2}-{\bf r}'')\nonumber \\
 &  & \times\langle{\bf h}({\bf r}')\cdot{\bf h}({\bf r}'')\rangle=\frac{h^{2}a^{3}}{\alpha_{e}^{2}}\int d^{3}rG_{k}({\bf r}-{\bf r}_{1})G_{k}({\bf r}-{\bf r}_{2})\nonumber \\
 &  & =\frac{h^{2}a^{3}}{\alpha_{e}^{2}}\int\frac{d^{3}q}{(2\pi)^{3}}\frac{e^{i{\bf q}\cdot({\bf r}_{1}-{\bf r}_{2})}}{(q^{2}+k^{2})^{2}}=\frac{h^{2}a^{3}}{8\pi\alpha_{e}^{2}k}e^{-k|{\bf r}_{1}-{\bf r}_{2}|}.
\end{eqnarray}
Writing the correlation function in the form $\langle{\bf S}({\bf r}_{1})\cdot{\bf S}({\bf r}_{2})\rangle=S_{0}^{2}\exp(-|{\bf r}_{1}-{\bf r}_{2}|/R_{f})$,
we finally obtain $k=1/R_{f}$,
\begin{equation}
\langle{\bf s}_{i}\cdot{\bf s}_{j}\rangle=s^{2}\exp\left(-\frac{|{\bf r}_{i}-{\bf r}_{j}|}{R_{f}}\right)\label{expCF}
\end{equation}
where
\begin{equation}
\frac{R_{f}}{a}=\frac{8\pi\alpha_{e}^{2}S_{0}^{2}}{h^{2}a^{4}}=8\pi\left(\frac{Js}{h}\right)^{2}.
\end{equation}
This result is in agreement with Eqs.\ (\ref{Rf-n}) and (\ref{CF-SR})
at $n=\infty$.

Two effects contribute to the energy associated with the disorder.
The first, that we call the short-range (SR) energy, is associated
with local disturbance of the ferromagnetic order by the RF. It does
not depend on the large-scale rotation of the magnetization characterized
by $R_{f}\gg a$. The second, that we call the long-range (LR) energy,
is associated with that large-scale rotation. Substituting Eq.\ (\ref{eq:S_via_h_mean_spherical})
into Eq.\ (\ref{H-continuous}) and integrating by parts we obtain
\begin{eqnarray}
 &  & {\cal H}=-\frac{1}{2\alpha_{e}}\int d^{3}rd^{3}r'd^{3}r''G_{k}({\bf r}-{\bf r}')\nabla_{r}^{2}G_{k}({\bf r}-{\bf r}'')\nonumber \\
 &  & \times{\bf h}({\bf r}')\cdot{\bf h}({\bf r}'')+\frac{1}{\alpha_{e}}\int d^{3}rd^{3}r'G_{k}({\bf r}-{\bf r}'){\bf h}({\bf r})\cdot{\bf h}({\bf r}').\nonumber \\
\end{eqnarray}
Here the first term originates from the exchange interaction and the
second term originates from the interaction of ${\bf S}$ with the
RF. Averaging this expression with the help of Eq.\ (\ref{eq:h-corr_continuous})
gives for the energy density
\begin{equation}
\frac{\langle{\cal H}\rangle}{V}=\frac{h^{2}a^{3}}{\alpha_{e}}\int d^{3}r\left[-\frac{1}{2}G_{k}({\bf r})\nabla^{2}G_{k}({\bf r})+G_{k}({\bf r})\delta({\bf r})\right]
\end{equation}
Recalling that $(\nabla^{2}-k^{2})G_{k}({\bf r})=\delta({\bf r})$,
for the energy per spin $E=(\langle{\cal H}\rangle/V)a^{3}$ one obtains
\begin{equation}
E=\frac{h^{2}a^{6}}{\alpha_{e}}\int d^{3}r\left[-\frac{k^{2}}{2}G_{k}^{2}({\bf r})+\frac{1}{2}G_{k}({\bf r})\delta({\bf r})\right].\label{E-density}
\end{equation}
Computation of the first term in Eq. (\ref{E-density}) involves integration
over distances of order $R_{f}$. It gives the long-range (LR) energy
of the mean spherical model ($n\rightarrow\infty)$:
\begin{equation}
E_{\mathrm{LR}}=-\frac{h^{4}}{128\pi^{2}J^{3}s^{2}}.\label{ELR}
\end{equation}
The last negative ($G_{k}<0$) term in Eq. (\ref{E-density}) that
contains the $\delta$-function is the combined short-range (SR) exchange
and RF energy, the RF energy being negative and twice the exchange
energy, both proportional to $h^{2}$. These integrals formally diverge
because of the singularity of $G_{k}(r)$ at $r=0$, which is an artifact
of the continuous field theory. They can be estimated by replacing
$G_{k}$ with $1/a$ at $r\rightarrow0$. It is possible, however,
to compute the SR energy approximately for the general $n$-component
model by using the discrete Hamiltonian, Eq. (\ref{eq:ham-discrete})
and the relation between the exchange energy and the nearest-neighbor
spin correlation function. Subtracting the energy $E_{0}=-3Js^{2}$,
of the ferromagnetically ordered state, one obtains for the SR exchange
energy per spin
\begin{equation}
E_{\mathrm{SR},ex}=3J[s^{2}-\langle{\bf s}(0)\cdot{\bf s}(a)\rangle]=\frac{3Js^{2}a}{R_{f}}=\left(1-\frac{1}{n}\right)\frac{3h^{2}}{8\pi J},
\end{equation}
where Eqs.\ (\ref{CF-SR}) and (\ref{Rf-n}) have been used. According
to the comment below Eq.\ (\ref{ELR}), the total SR energy is
\begin{equation}
E_{\mathrm{SR}}=-\left(1-\frac{1}{n}\right)\frac{3h^{2}}{8\pi J}.\label{E-SR}
\end{equation}

\section{Numerical method}

\label{sec:numerical}

Random-field systems exhibit glassy behavior with many local energy
minima. Same as in Ref. \cite{GCP-PRB13}, at $T=0$ we minimize
the energy of the system in the process of the relaxation from a typical
initial state such as, e.g., random or collinear orientation of spins.
The method consists of the sequential rotation of each spin towards
the direction of the effective field at a given site with the probability
$\alpha$ or its flipping over the effective field (the so-called
over-relaxation, conserving the energy) with the probability $1-\alpha$.
Computation based upon rotations only had been developed for random-anisotropy
systems in Ref. \cite{Dieny-PRB1990}. The addition of the
flipping (the so-called over-relaxation) that conserves energy makes
the method more effective. The parameter $\alpha$ has the meaning
of the relaxation constant. For our glassy systems the method works
most efficiently when small $\alpha$ is used. Most of the results
have been obtained with $\alpha=0.03$. The resulting energy minima
are representative local energy minima. No attempt has been made to
find the global energy minimum as it would be hopeless for the glassy
system of a large size that we have studied. In the physical applications,
high energy barriers prevent the glassy system from reaching the global
energy minimum.

At nonzero temperatures we use Monte Carlo method described in Sec.
\ref{sub:MC_method}.

In the numerical work we consider a simple cubic lattice with periodic
boundary conditions and set $s=a=J=1$, using $H_{R}$ instead of
$h$. The numerical method has been implemented in Wolfram Mathematica
in a compiled and parallelized form. To reduce statistical fluctuations
in our random system, one can do averaging over realizations of the
RF. For systems of large sizes that we use the self-averaging over
the system produces the same effect. To avoid large fluctuations the
system size $L$ must satisfy $R_{f}\ll L$. For the weak random field,
Eq. (\ref{Rf-n}) gives large $R_{f}$ so that a large $L$ is needed
to prove ordering or disordering. For $R_{f}\gtrsim L$ the system
has a large finite-size magnetization in all cases.

Vorticity in the $xy$ model can be computed by considering rotations
of spins when one circles the lattice site in the positive direction
(counterclockwise) around each square plaquette within $xy$, $yz$,
and $zx$ planes. \cite{Gingras-Huse-PRB1996,GCP-PRB13}. If the spins
rotate by a zero angle, there is no singularity. Rotation by $\pm2\pi$
corresponds to the vortex/antivortex. For models with arbitrary number
of spin components for each cube of neighboring spins the average
spin vector and its length can be computed. If the latter is smaller
than a preset value (we used 0.5), there is a singularity at this
point.
In particular, in the $3d$ Heisenberg model there are so-called hedgehogs,
singularity points around which the magnetization points towards the center or away from the center.
The proposed method reliably provides positions of hedgehogs in the lattice.
The fraction of positions having singularities is reported
as $f_{S}$. This method works well for any dimensionality of space
$d$ and any number of spin components $n$. For the $xy$ model it
yields the results consistent with those obtained by looking for vortices/antivortices.

\section{Energy and magnetization}

\label{E vs m}

\begin{figure}
\includegraphics[width=8cm]{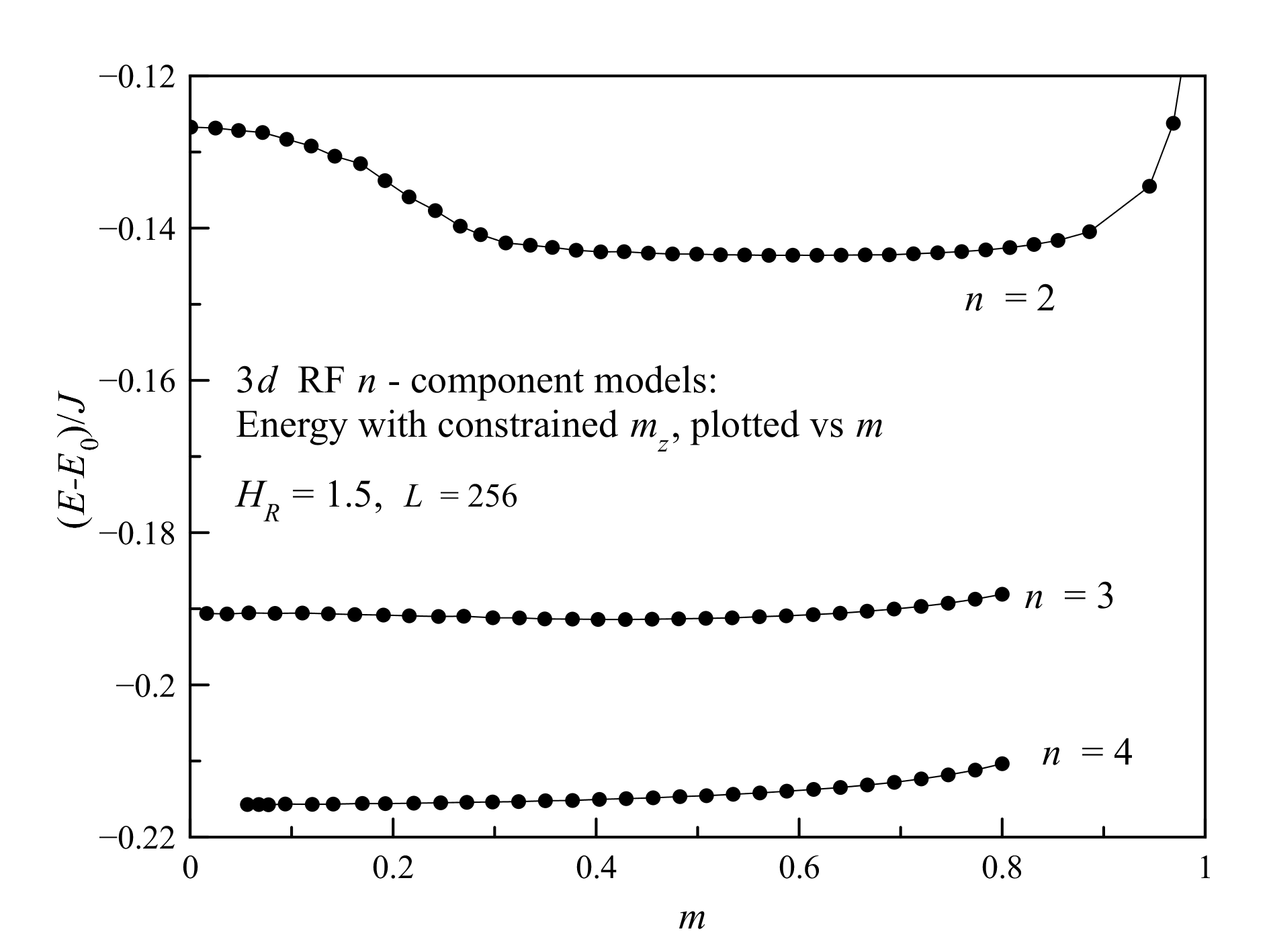}\caption{Energy $E$ vs the magnetization $m$ for different numbers of spin
components, $n=2,3,4$. Each point was obtained by relaxation from a random
initial state.}

\label{Fig-dEconstr_vs_m_Nalp=2_3_4_L=256_HR=1.5_pbc_alp=0.03_rand_IC}
\end{figure}

\begin{figure}
\includegraphics[width=8cm]{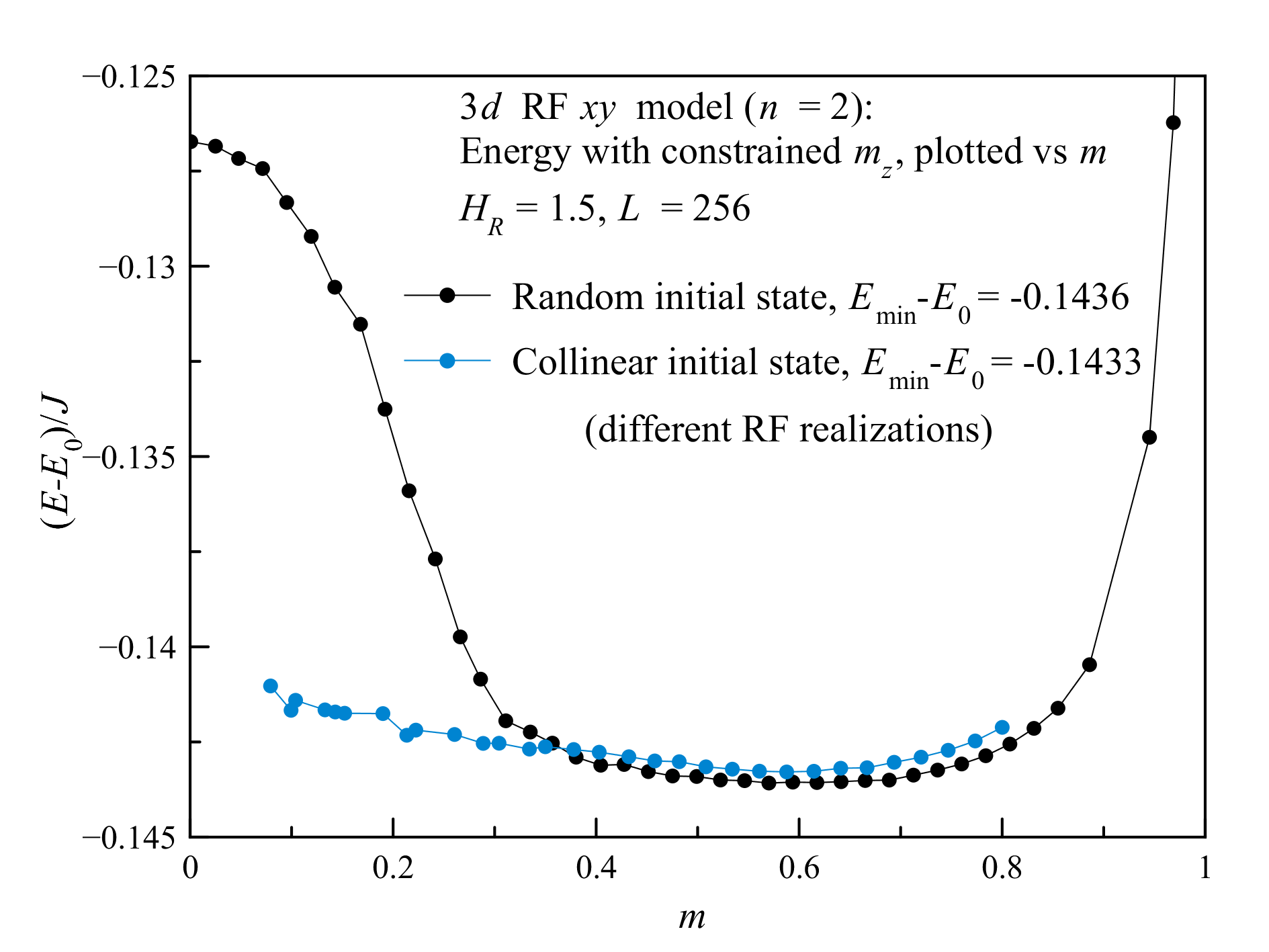}

\includegraphics[width=8cm]{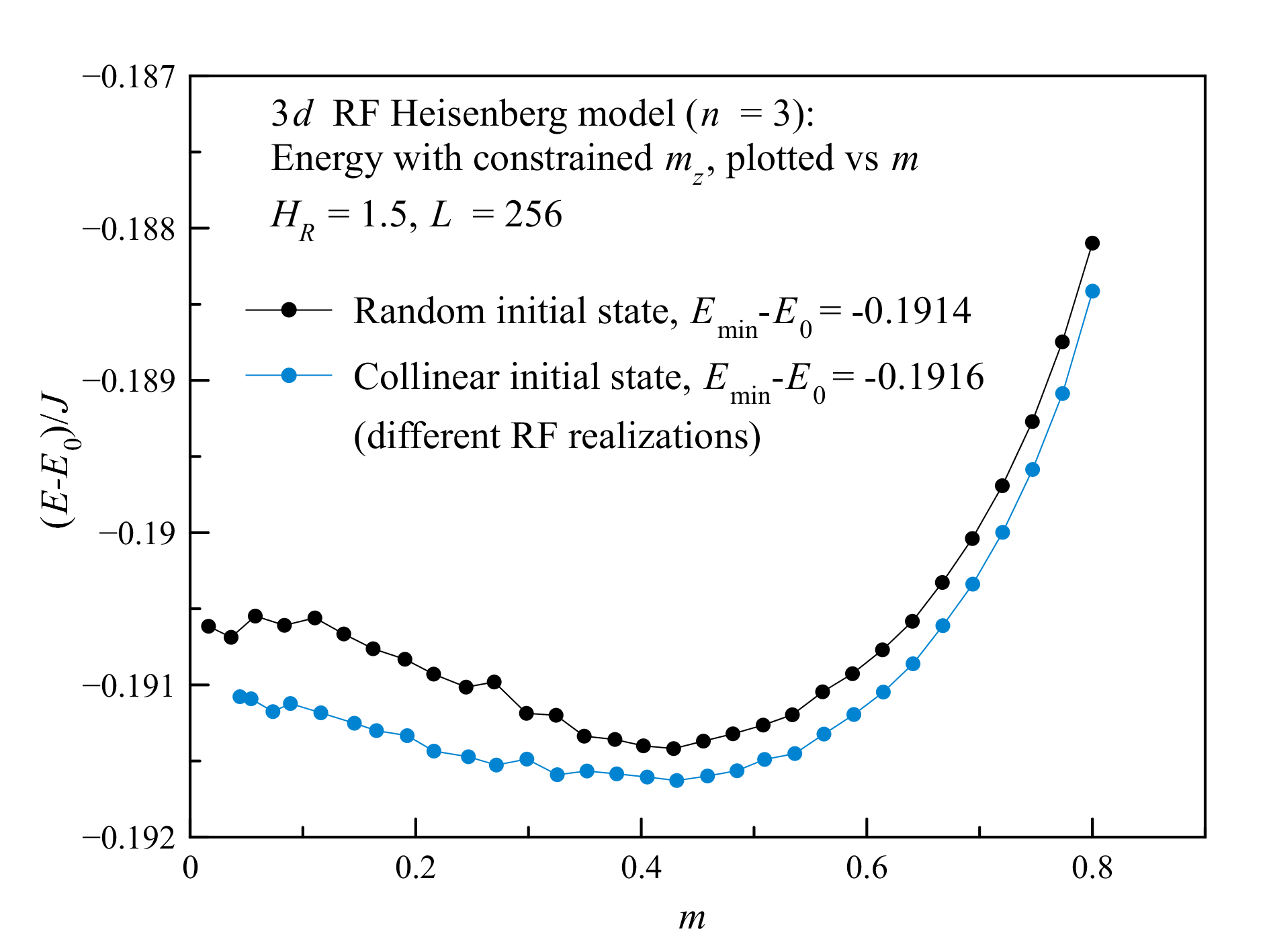}

\includegraphics[width=8cm]{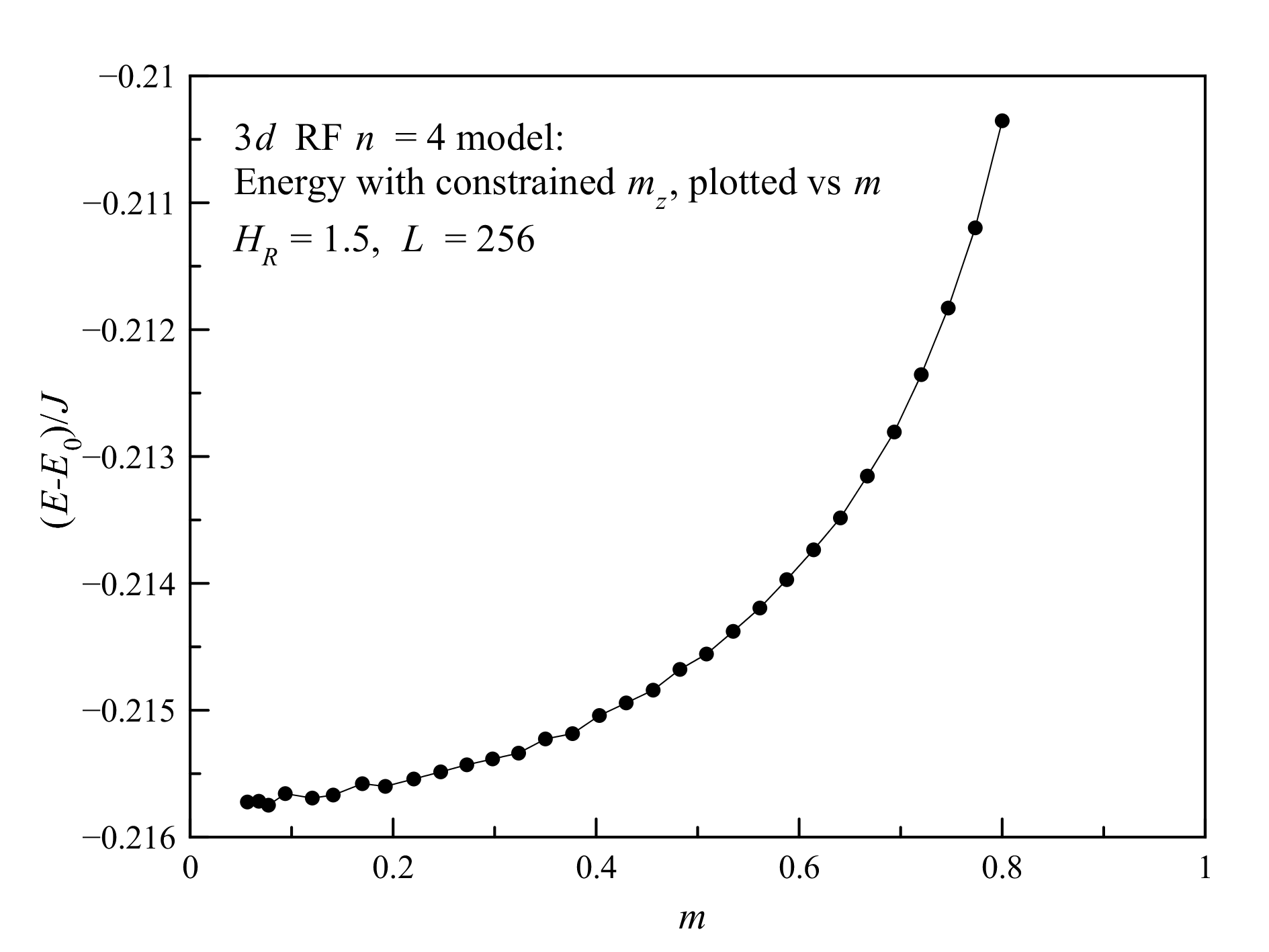}

\caption{Energy $E$ vs the magnetization $m$ for different models. Each point
was obtained by relaxation from a random or collinear initial state. Note that here the energy scale is finer than in Fig.\ \ref{Fig-dEconstr_vs_m_Nalp=2_3_4_L=256_HR=1.5_pbc_alp=0.03_rand_IC}.}

\label{Fig-dEconstr_vs_m_L=256_HR=1.5_pbc_alp=0.03_rand_IC}
\end{figure}

Correlation of the energy of local energy minima with the magnetization
in the $xy$ model has been studied in Ref. \cite{GCP-PRB13}.
The energy of the vortex-glass (VG) state was found to correlate perfectly
with the vorticity and to be higher than the energy of the F-state.
The latter had a flat minimum at $m$ between 0.5 and 0.6 for $H_{R}/J=1.5$.
The F-states with lower magnetization (the lowest being just below
$m=0.2$) have manifestly higher energies.

Here a different computation has been performed for models with different
$n$. The energy was minimized under the constraint of $m_{z}=\mathrm{const}$
applying a self-adjusting field $H$ along the $z$ axis as a Lagrange
multiplier. We measure energy with respect to the exchange energy,
$E_{0}$, of the collinear ferromagnetic state. The resulting difference,
$E-E_{0}$, (without the Lagrange Zeeman energy) was plotted vs resulting
magnetization $m$. Although in the computations the target value of $m_{z}$ was fixed,
other components of $\mathbf{m}$ were not, so that their
non-zero values contributed to $m$. The initial state for the relaxation
was random or collinear orientation of spins. The results for different
$n$ and random initial conditions are shown in
Fig. \ref{Fig-dEconstr_vs_m_Nalp=2_3_4_L=256_HR=1.5_pbc_alp=0.03_rand_IC},
while the results for different models separately are shown in a finer scale in
Fig. \ref{Fig-dEconstr_vs_m_L=256_HR=1.5_pbc_alp=0.03_rand_IC}.
The curve $E(m)$ for the $xy$ model ($n=2$) has a pronounced flat
minimum. Higher energy for $m\lesssim0.2$ is mainly due to vortex
loops. Higher energies for $m$ close to 1 are due to the near-collinearity
of spins and their non-aligning with the random field.

The curves
for $n=3,4$ are nearly flat in
Fig. \ref{Fig-dEconstr_vs_m_Nalp=2_3_4_L=256_HR=1.5_pbc_alp=0.03_rand_IC}.
However, one can see a minimum of the energy at a finite $m$ for the Heisenberg model ($n=3$)
in the middle panel of
Fig. \ref{Fig-dEconstr_vs_m_L=256_HR=1.5_pbc_alp=0.03_rand_IC}.
The latter occurs due to hedgehogs
in the region of small $m$ that inevitably occur due to topology \cite{PGC-PRL2014}. Their energy is much smaller than the
energy of vortex loops in the $xy$ model, thus the energy minimum is shallow relative to the $xy$ model.
The effect of hedgehogs is subtle.
In Sec. \ref{hysteresis} it will be shown that a more sophisticated method of finding the energy minimum with the help
of hysteresis loops with the amplitude of the magnetic field gradually decreasing to zero allows the system to better adjust
and attain the minimal-energy state with $m=0$ in spite of hedgehogs.

For $n=4$ there are no
singularities \cite{PGC-PRL2014}, and the position of the energy minimum can be projected
to $m=0$. Fig. \ref{Fig-dEconstr_vs_m_L=256_HR=1.5_pbc_alp=0.03_rand_IC}
shows all three curves separately. One can see the energy minimum
slightly above $m=0.4$ for the Heisenberg model, $n=3$.

The presence or absence of singularities in the completely disordered states of the random-field model follow from the
topological argument of Ref. \cite{PGC-PRL2014}. As pointed out by Imry and Ma \cite{Imry-Ma-PRL1975}, in the completely disordered state
the magnetization should follow the direction of the random field averaged over large areas $\bf \bar h$. Different components of the average random field
$\bar h_\beta$, $\beta=1,\ldots,n$ are statistically independent and take different signs. Equations $\bar h_\beta=0$, $\beta=1,\ldots,n$
define $n$ surfaces of dimension $d-1$ in the $d$-dimensional space of the system. For $n\leq d$ there are subspaces where  $\bf \bar h =0$
and the direction of the magnetization is undefined.
Crossing these subspaces cause the magnetization vector to abruptly change its direction \cite{PGC-PRL2014}, that is,
the subspaces $\bf \bar h =0$ correspond to singularities in the spin field.

\section{Evolution of the F-state during the hysteresis cycle}

\label{hysteresis}

The hysteresis loop in the $3d$ $xy$ model is rather wide because
of the formation of topologically stable 180$^\circ$ ``spin walls'' as the field changes its direction.
In spin walls spins are still directed in the old direction,
whereas everywhere else the direction of spins follow that of the field
(see Sec. "Hysteresis loops" of Ref. \cite{Dieny-PRB1990} and Sec. 5H of Ref. \cite{GCP-PRB13}).
Spin walls rupture at a particular
negative field, creating vortex loops. \cite{GCP-PRB13} In the $3d$
Heisenberg model the hysteresis loop is apparently due to hedgehogs
and is much narrower, while the hysteresis loop for the non-singular model with
$n=4$ it almost invisible. \cite{PGC-PRL2014}

\begin{figure}
\includegraphics[width=8cm]{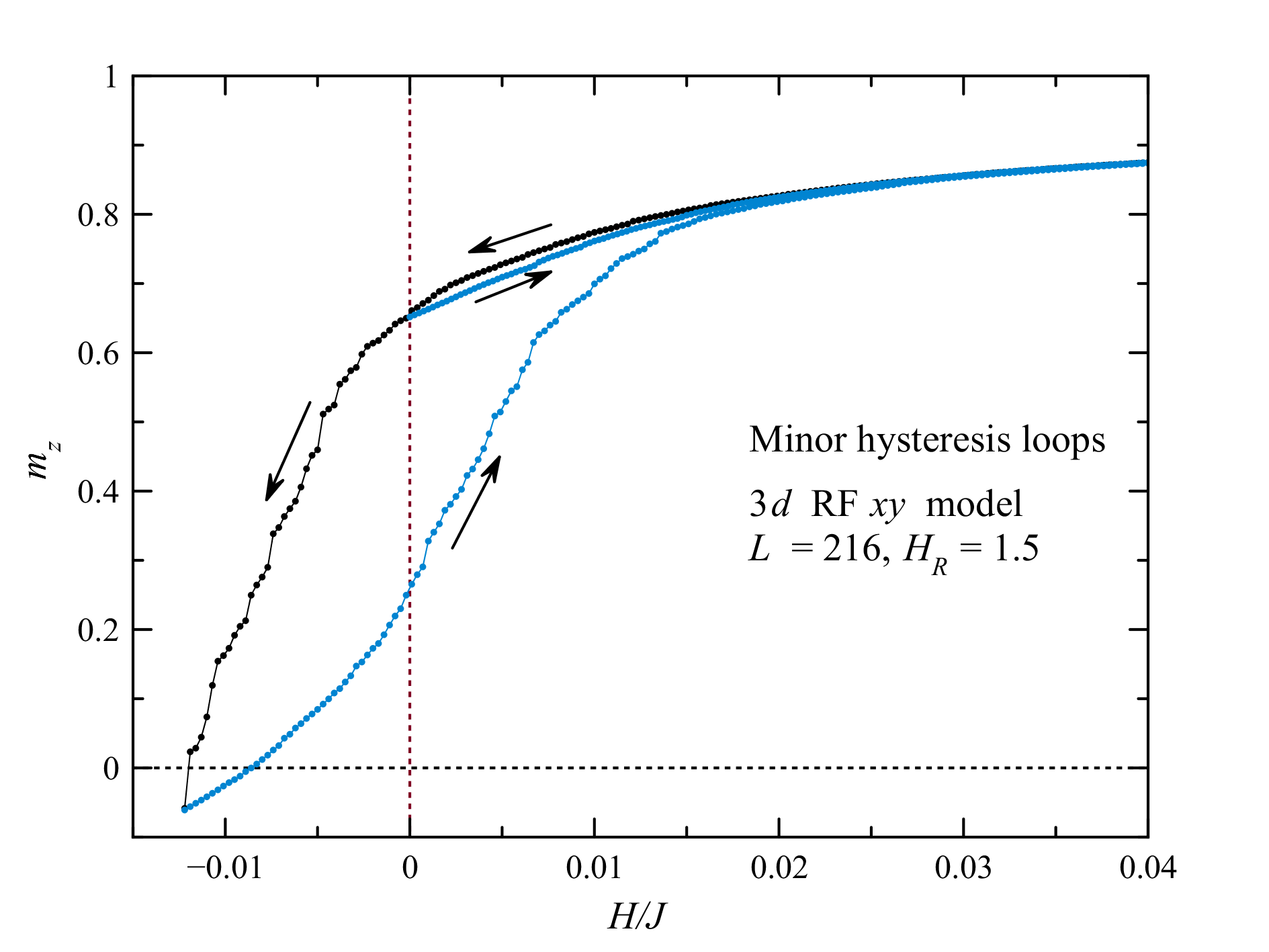}

\caption{Minor hysteresis loops in the $3d$ $xy$ model. All these states
are vortex-free.}

\label{Fig-mz_vs_H_Nalp=2_L=216_HR=1.5_pbc_alp=0.03-minor}
\end{figure}

\begin{figure}
\includegraphics[width=8cm]{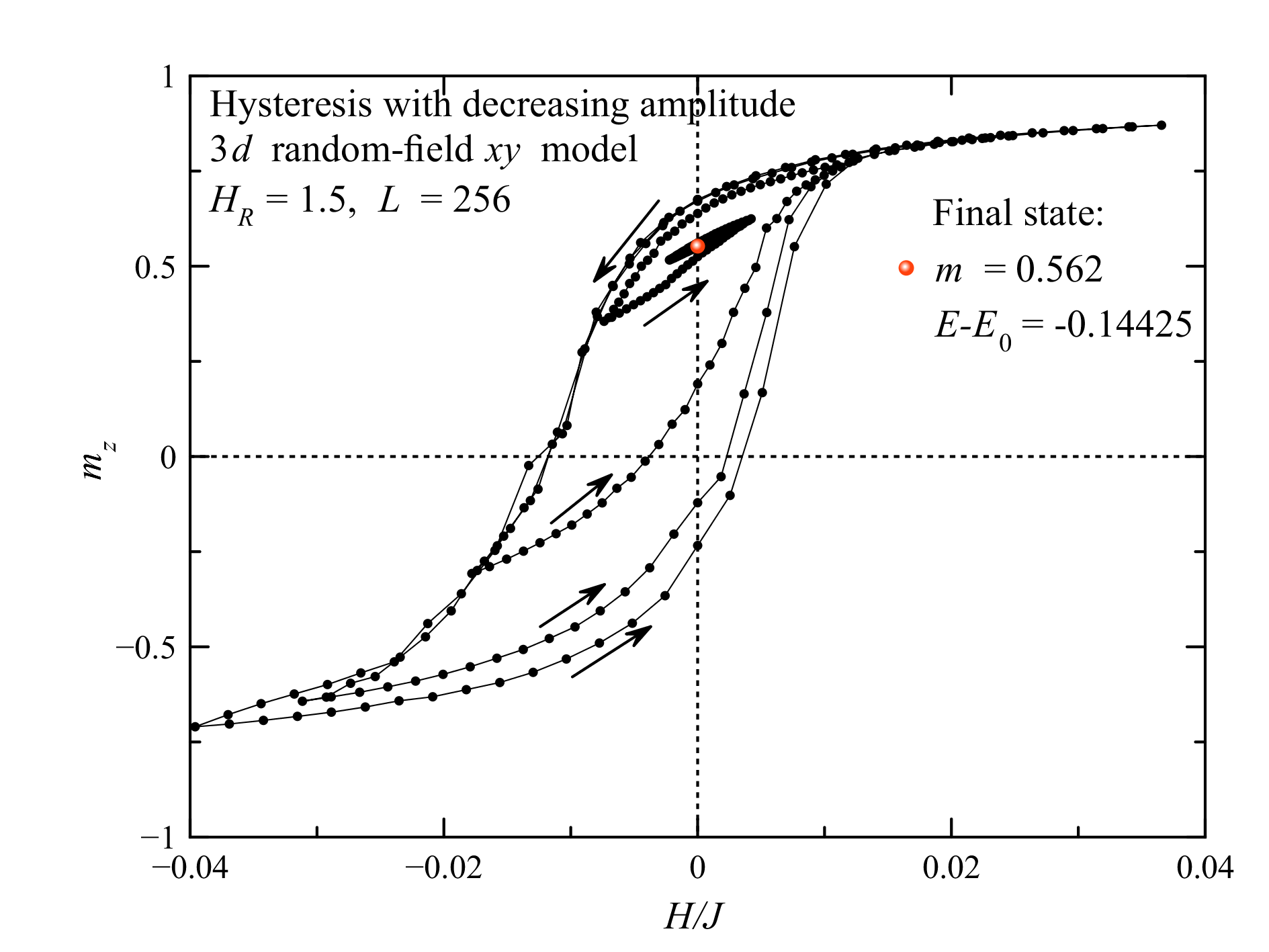}

\caption{Evolution of the magnetization in the $3d$ $xy$ model during the
hysteresis cycle with amplitude decreasing to zero, starting from positive fields. All states are
vortex free. Using larger initial amplitudes of the magnetic field leads to rupture of spin walls
at negative fields and thus creating vortex loops that increase the energy of the system.}

\label{Fig-mz_vs_H_decr_ampl_3d_n=2_L=256_HR=1.5}
\end{figure}

\begin{figure}
\includegraphics[width=8cm]{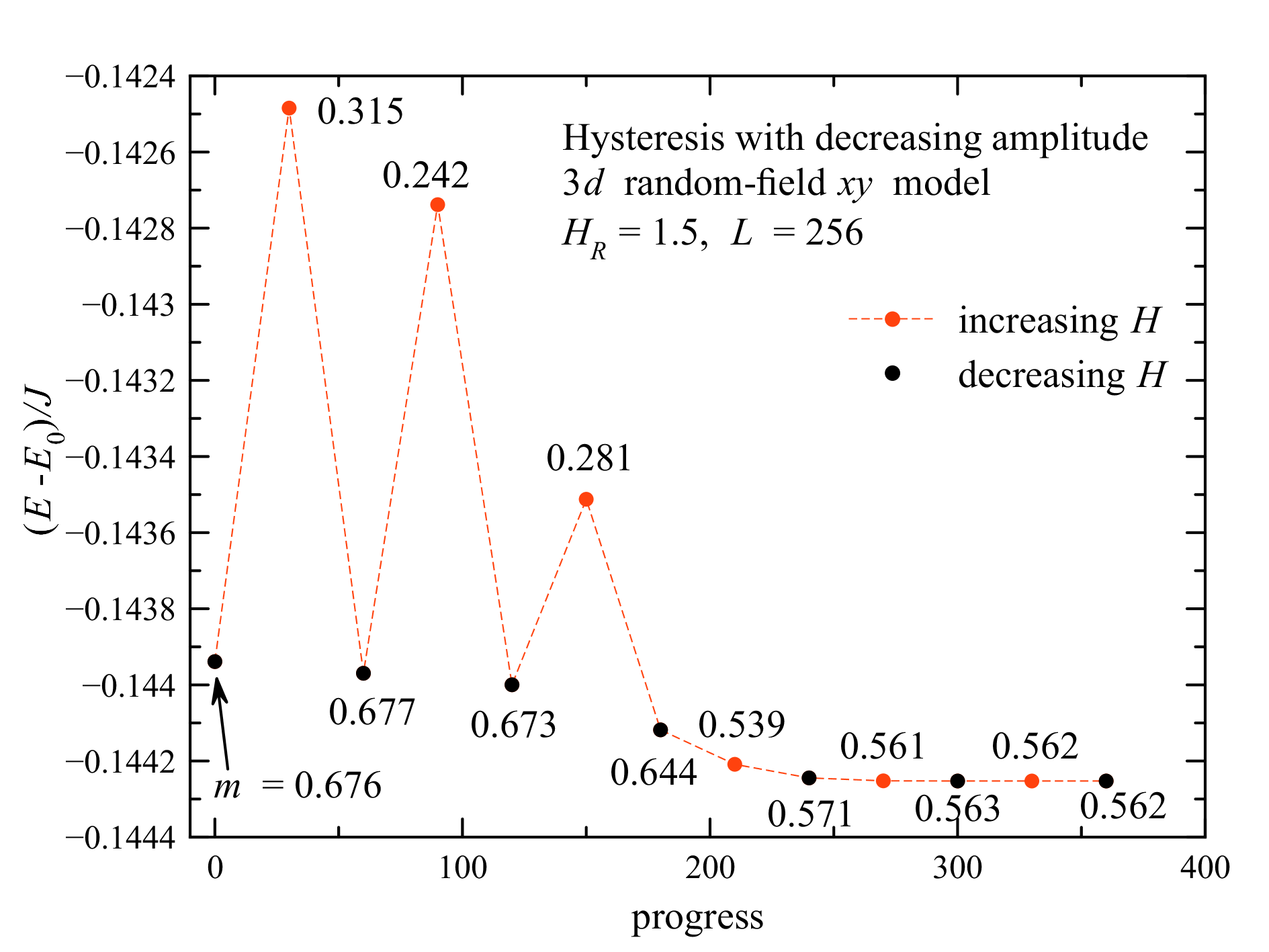}

\caption{Evolution of energy in the $3d$ $xy$ model during the hysteresis
cycle with decreasing amplitude. Only points corresponding to $H=0$
are shown. Black (red) circles correspond to decreasing (increasing)
$H$.}

\label{Fig-dE_vs_progress_3d_n=2_L=256_HR=1.5}
\end{figure}

\begin{figure}
\includegraphics[width=8cm]{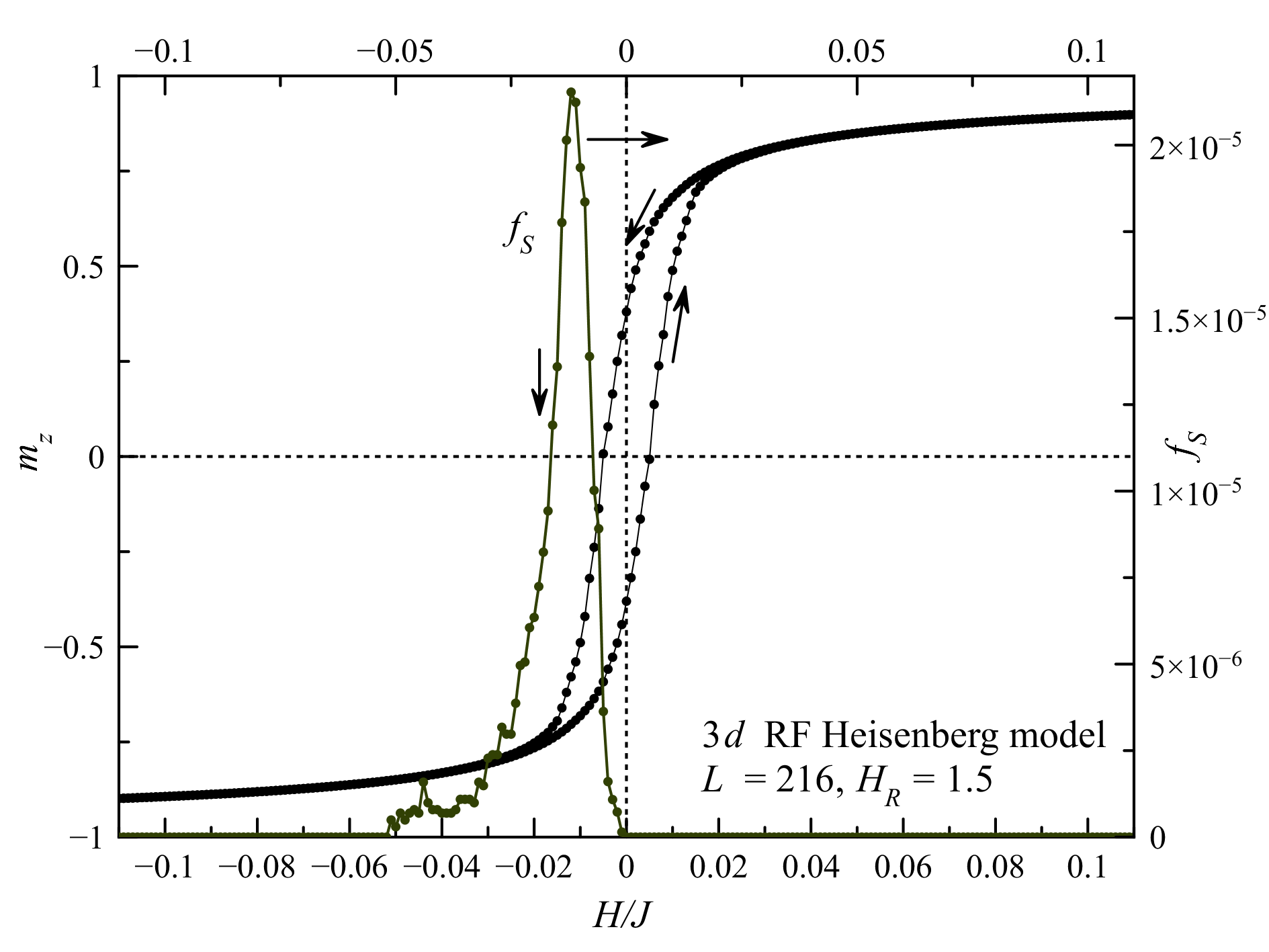}

\caption{Hysteresis loop of the $3d$ Heisenberg model.}

\label{Fig-mz_vs_H_Nalp=3_Nx=Ny=Nz=216_HR=1.5}
\end{figure}

\begin{figure}
\includegraphics[height=6cm]{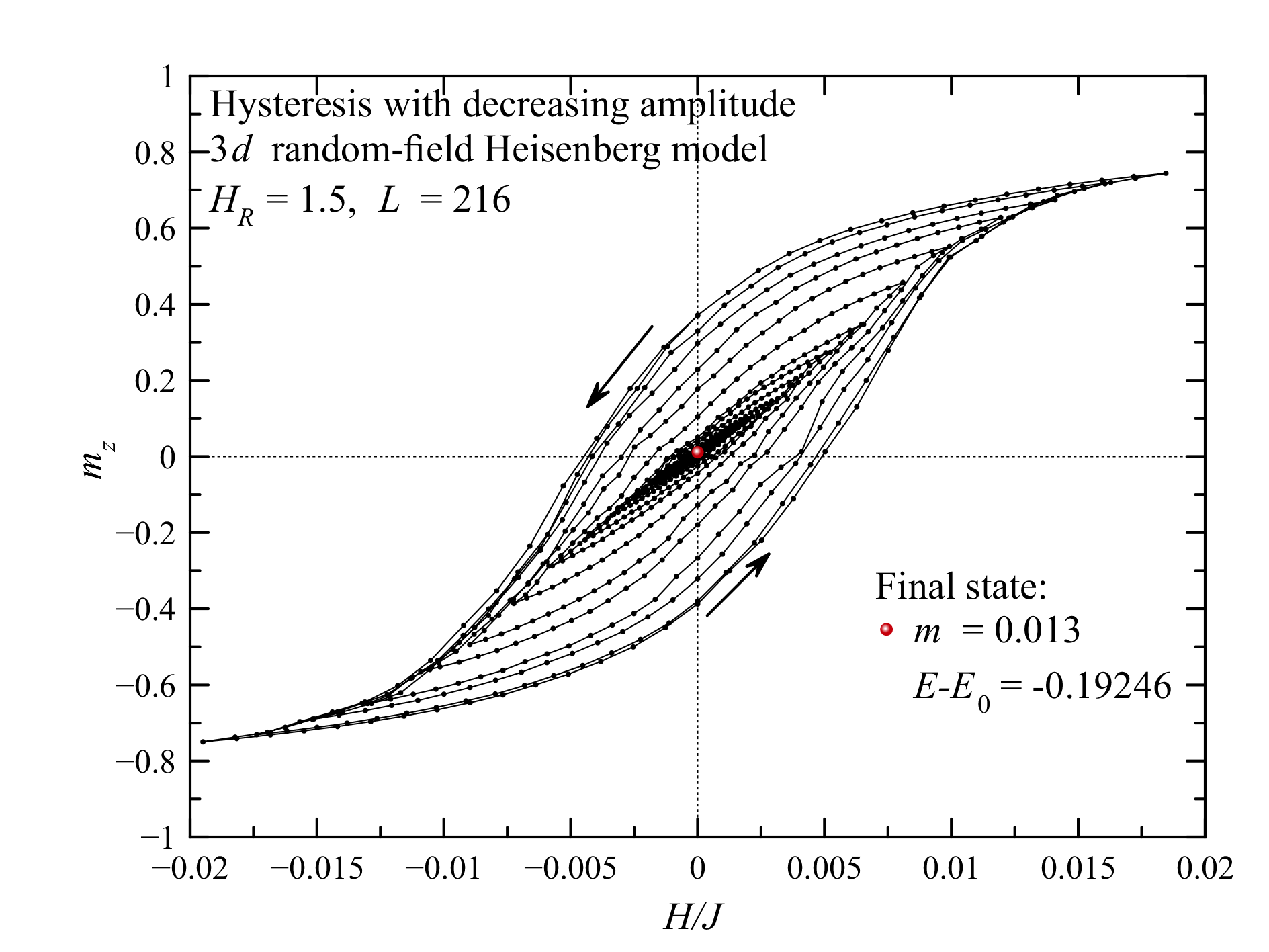}

\caption{Evolution of the magnetization in the $3d$ Heisenberg model during
the hysteresis cycle with amplitude decreasing to zero.}

\label{Fig-mz_vs_H_decr_ampl_3d_n=3_L=216_HR=1.5}
\end{figure}

\begin{figure}
\includegraphics[height=6cm]{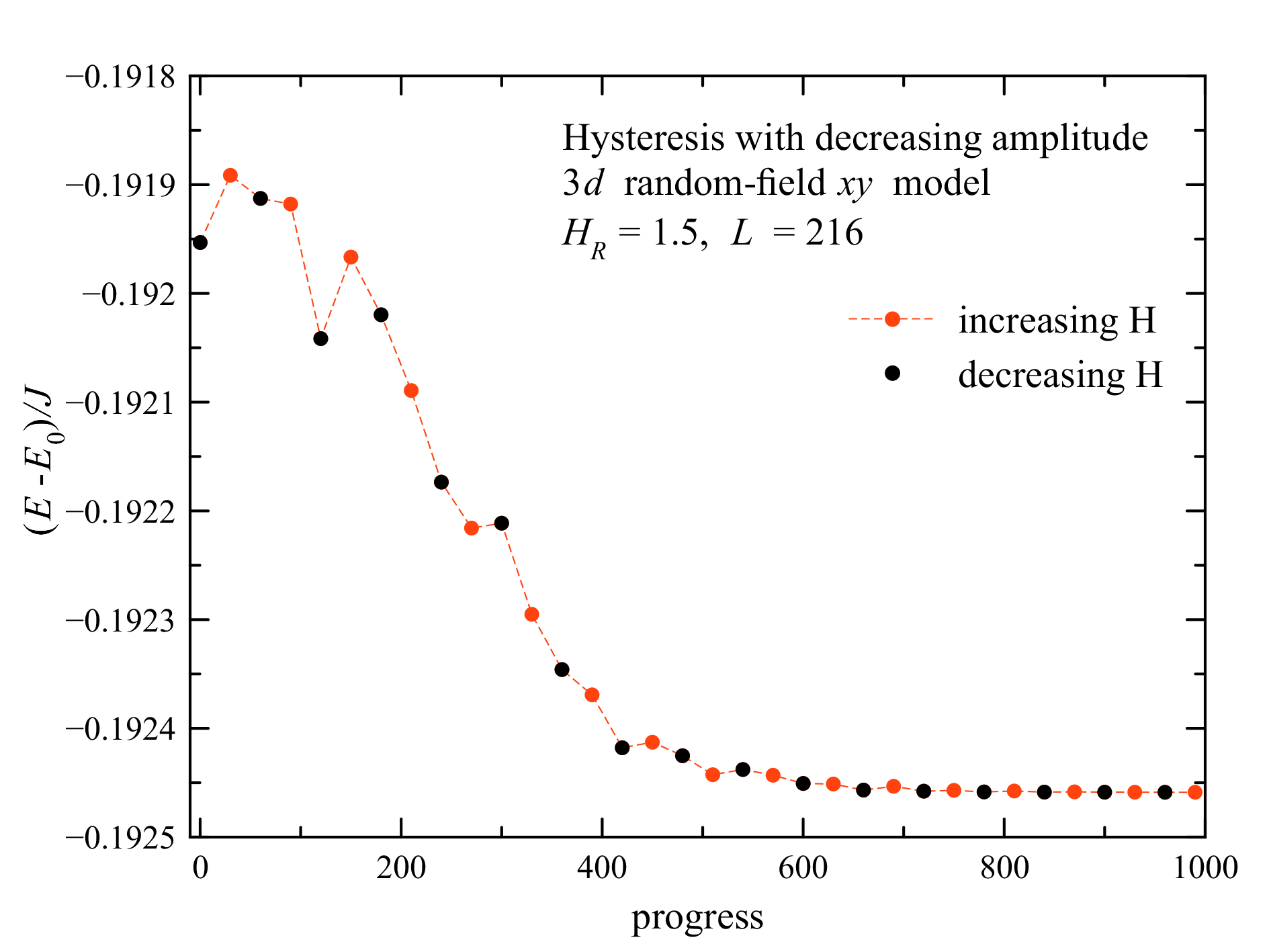}

\caption{Evolution of energy in the $3d$ Heisenberg model during the hysteresis
cycle with decreasing amplitude. Only points corresponding to $H=0$
are shown. Black (red) circle correspond to the decreasing (increasing)
$H$.}

\label{Fig-dE_vs_progress_3d_n=3_L=216_HR=1.5}
\end{figure}

\begin{figure}
\includegraphics[height=6cm]{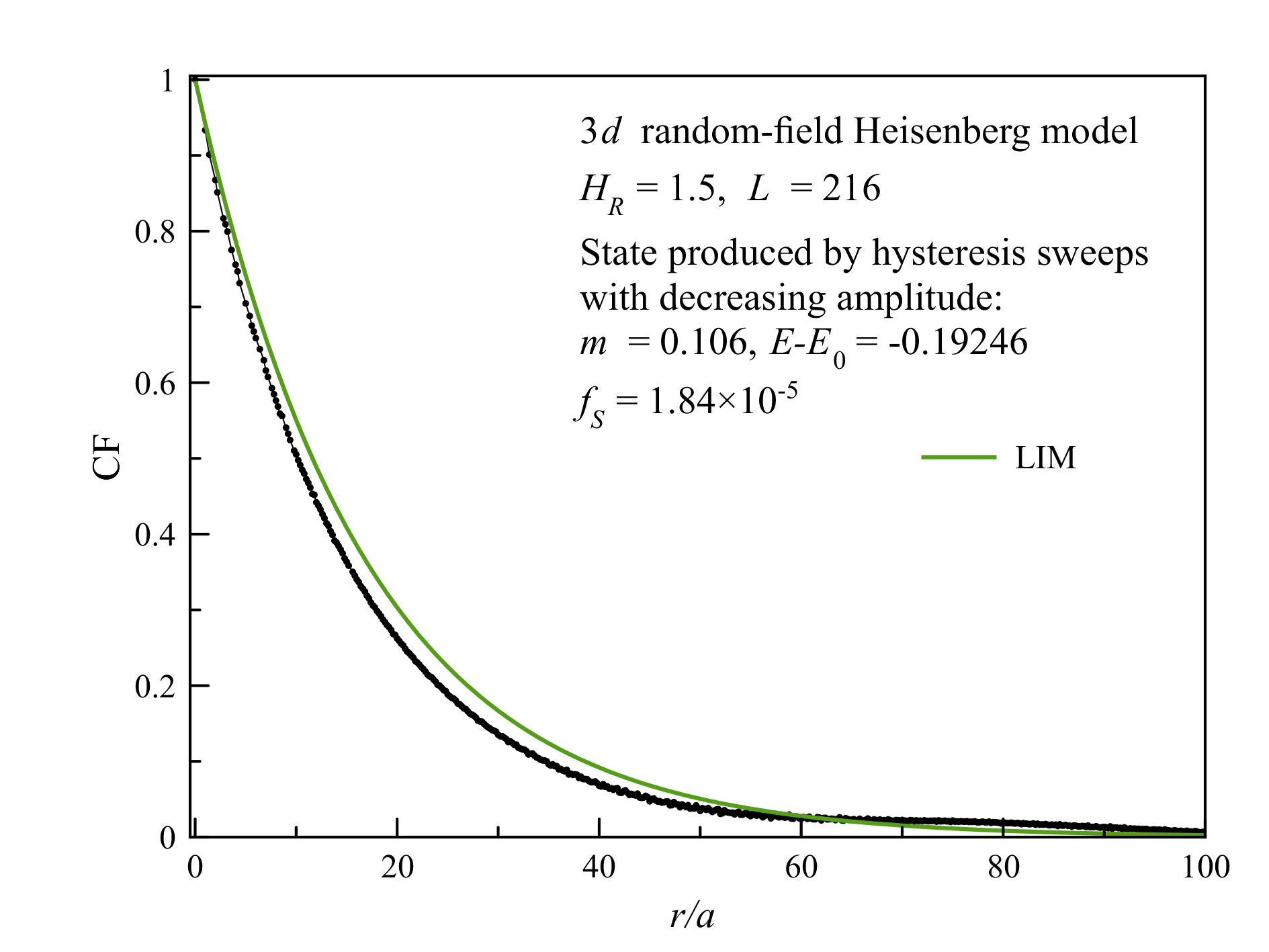}

\caption{Spin-spin correlation function of the $3d$ Heisenberg model at the
end of the hysteresis cycle with decreasing amplitude. }

\label{Fig-CF_vs_n_decreasing_amplitude_3d_n=3_L=216_HR=1.5}
\end{figure}

Here we show that the glassy properties of the RF models are not exclusively
due to the singularities and that barriers also exist in the vortex-free
states of the $xy$ model, leading to minor hysteresis loops. Such
loops are shown in Fig. \ref{Fig-mz_vs_H_Nalp=2_L=216_HR=1.5_pbc_alp=0.03-minor}.
Making repeated hysteresis loops with the amplitude of the applied
field $H$ decreasing to zero, one can hope to end up in a state of
a smaller energy. To save the computation time, the process can be
started as free relaxation in zero field from a collinear state aligned
with the direction in which $H$ will be applied. This results in
the F-state. Then the field is increased in the direction opposite
to the initial magnetization etc. For the $3d$ $xy$ model, starting
with a large $m$ leads to the rupture of spin walls and creation
of the vortex loops. As the result, the system ends up in a state
of a higher energy than the initial F-state.

For the $3d$ $xy$ model one can lower the energy of the initial
F-state by making hysteresis loops with the initial amplitude smaller
than the magnetic field that causes rupture of spin walls. The evolution
of the magnetization in this sequence vs the applied field is shown
in Fig. \ref{Fig-mz_vs_H_decr_ampl_3d_n=2_L=256_HR=1.5}.
Evolution of the energy is shown in Fig. \ref{Fig-dE_vs_progress_3d_n=2_L=256_HR=1.5}
(zero values of $H$ only). One can see that the final energy is indeed
lower than the initial energy $\left(E-E_{0}\right)/J=-0.1439$ of
the state with $m=0.676$ obtained by relaxation from the collinear
state (the first point in Fig. \ref{Fig-dE_vs_progress_3d_n=2_L=256_HR=1.5}).
For $H_{R}/J=1.5$ and $L=256$ the final state has $m=0.562$ and
$\left(E-E_{0}\right)/J=-0.14425$. Another computation with $H_{R}/J=1.5$
and $L=216$ resulted in $m=0.567$ and $\left(E-E_{0}\right)/J=-0.1441$.
It is important to notice that the amplitude of the hysteresis loop
remains small enough so that rupture of spin walls accompanied by
the creation of vortex loops \cite{GCP-PRB13} does not occur. Otherwise
the process would lead to the energy increase: the final state (also
having the large magnetization typical for the F-state) contains singularities
and its energy is higher than that of the F-state obtained by relaxation
from the collinear state. These experiments show that the F-state
of the $3d$ $xy$ model is robust. It cannot be easily destroyed
by manipulating the external field.

A completely different behavior has been observed for the $3d$ Heisenberg
model. Here, as shown in Fig. \ref{Fig-mz_vs_H_decr_ampl_3d_n=3_L=216_HR=1.5},
decreasing the magnetic field to negative values immediately leads
to the formation of singularities - hedgehogs. Thus hedgehogs are
unavoidable and one can start with a greater amplitude of the field
sweep, eventually reducing it to zero, which leads to $m=0.013$, as shown in
Fig. \ref{Fig-mz_vs_H_decr_ampl_3d_n=3_L=216_HR=1.5}.
Such a small magnetization can be explained by the finite-size effect
in the absence of ordering. Disappearance of the magnetization is
accompanied by the decrease of the energy in Fig. \ref{Fig-dE_vs_progress_3d_n=3_L=216_HR=1.5}
down to $\left(E-E_{0}\right)/J=-0.19246$. The fraction of singularities
in the final state is $f_{S}=1.84\times10^{-5}$. One can see that
for the $3d$ Heisenberg model the energy increase due to the creation
of singularities as spins align with the RF is smaller than the energy
gain due to the aligning. Although relaxation from the collinear initial
state stops at a finite $m$ because of the singularities
(see the middle panel of Fig. \ref{Fig-dEconstr_vs_m_L=256_HR=1.5_pbc_alp=0.03_rand_IC}), a more
sophisticated process of the magnetic field sweep with decreasing
amplitude helps the system to attain lower energy by disordering completely.
The spin-spin correlation function in the final disordered state shown
in Fig. \ref{Fig-CF_vs_n_decreasing_amplitude_3d_n=3_L=216_HR=1.5}
is close to the LIM exponential form. Slightly lower correlations
in the numerical result can be explained by hedgehogs.

\begin{figure}
\includegraphics[width=8cm]{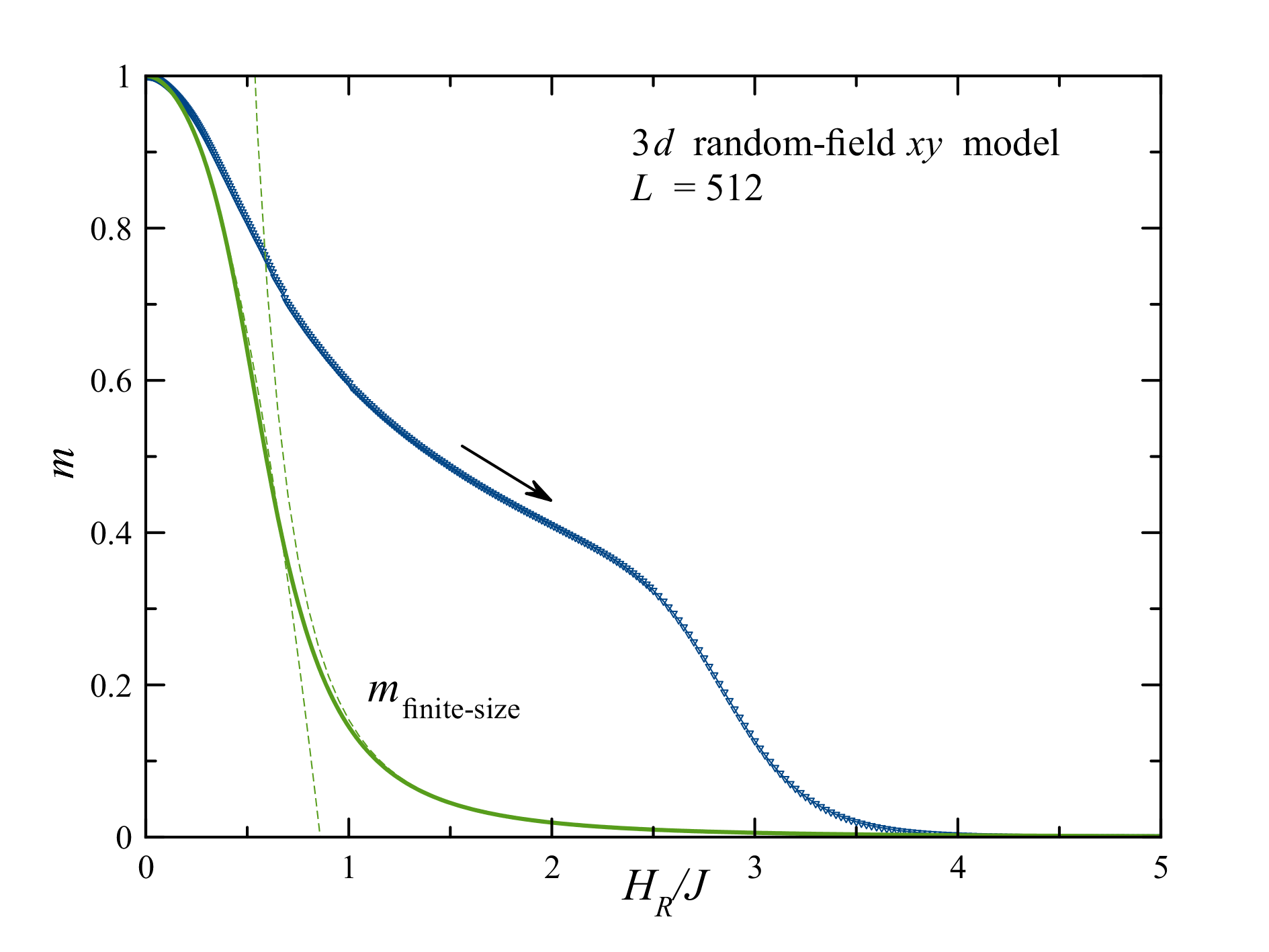}

\caption{Magnetization vs the RF strength $H_{R}$ for the $3d$ $xy$ model,
$L=512$. Green line is the finite-size magnetization expected in
the absence of ordering. Dashed green lines are given by Eqs. (\ref{eq:m-small_L-result})
and (\ref{eq:m-large_L-result}).}

\label{Fig-m_vs_HR_L=512}
\end{figure}

\begin{figure}
\includegraphics[width=8cm]{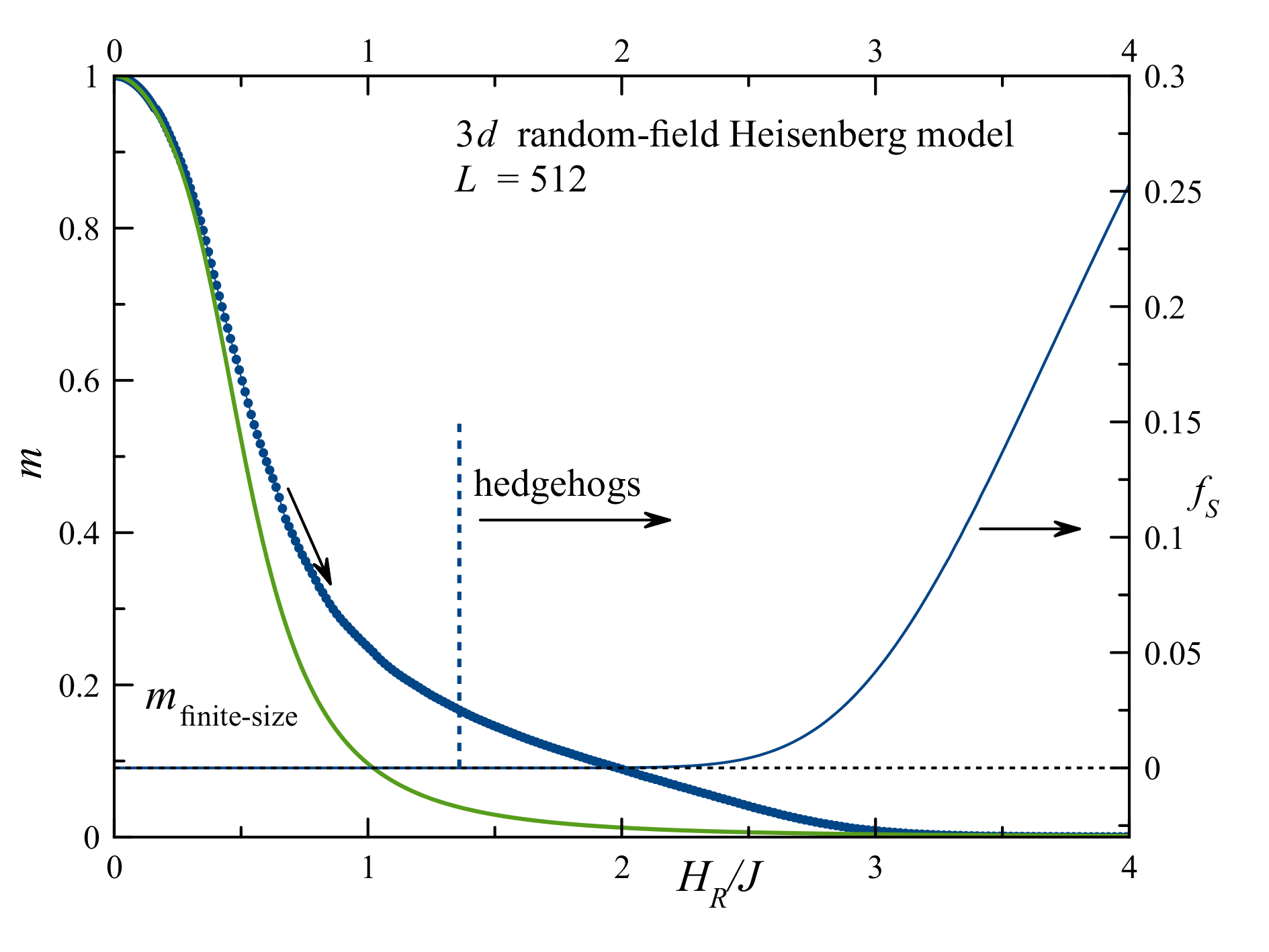}

\caption{Magnetization vs the RF strength $H_{R}$ for the Heisenberg model,
$L=512$. Green line is the finite-size magnetization expected in
the absence of ordering.}

\label{Fig-m_vs_HR_n=3_L=512_pbc}
\end{figure}

\begin{figure}
\includegraphics[width=8cm]{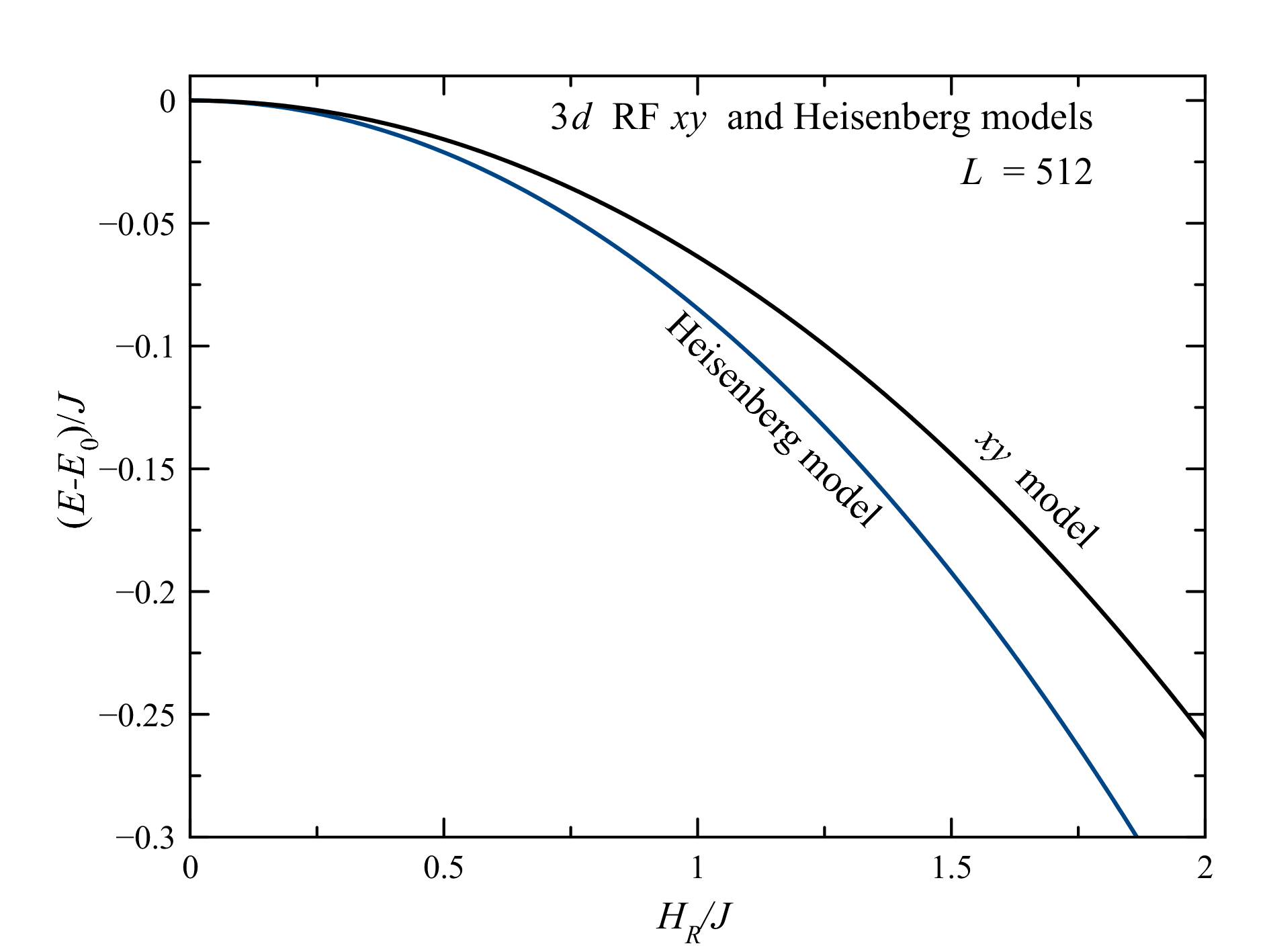}\caption{Energy vs the RF strength $H_{R}$ for the $xy$ and Heisenberg models,
$L=512$.}

\label{Fig-dE_vs_HR_n=2_3_L=512}
\end{figure}

\section{Magnetization of the F-state vs random-field strength}

\label{m vs h}

Another way to obtain the F-states is, starting from a collinear state,
to increase the strength of the random field $H_{R}$ from zero in
small steps. \cite{GCP-PRB13} This procedure leads to the F-states
with a smaller magnetization than that obtained directly by the relaxation
from a collinear state. Results for magnetization and energy for the
$3d$ $xy$ model at $H_{R}/J=1.5$ are $m=0.467$ and $\left(E-E_{0}\right)/J=-0.14425$.
New computation for $H_{R}/J=1.5$ and $L=512$, Figs. \ref{Fig-m_vs_HR_L=512},
\ref{Fig-m_vs_HR_n=3_L=512_pbc},
and \ref{Fig-dE_vs_HR_n=2_3_L=512},
yields $m=0.486$ and $\left(E-E_{0}\right)/J=-0.14415$ for the $xy$
model and $m=0.146$ and $\left(E-E_{0}\right)/J=-0.19226$ for the
Heisenberg model. Comparison of these results with those above shows
that the energy landscape of the RF model is very flat and states
with markedly different magnetizations have very close energies.

The computed magnetization should be compared with the finite-size
magnetization expected in the absence of ordering in the intermediate
range of $H_{R}$. The latter can be obtained analytically in the
regions $R_{f}\gg L$ and $R_{f}\ll L$, where the magnetic correlation
range $R_{f}$ is given by Eq.\ (\ref{Rf-n}) and $h\equiv H_{R}$.
For small $H_{R}$ one has $R_{f}\gg L$ and the short-range order
extends over the whole system, making the magnetization $m\cong1$.
Deviations from this short-ordered state can be found perturbatively.
For the cube of size $L$ with periodic boundary conditions the result
has the form
\begin{equation}
m\cong1-A\frac{L}{R_{f}}\cong\left(1+A\frac{L}{R_{f}}\right)^{-1},\label{eq:m-small_L-result}
\end{equation}
where
\begin{equation}
A=\frac{1}{4\pi^{3}}\sum_{n_{x,y,z}=-\infty}^{\infty}\!^{'}\frac{1}{\left(n_{x}^{2}+n_{y}^{2}+n_{z}^{2}\right)^{2}}\simeq0.1333\cong\frac{4}{30}.
\end{equation}
In the case $R_{f}\ll L$, assuming the exponential spin correlation
function of Eq.\ (\ref{expCF}) (that has been derived analytically
for $n=\infty$ in Section \ref{analytical} and confirmed numerically
\cite{PGC-PRL2014} for $3d$ models with $n>d+1$) one obtains\cite{GCP-PRB13}
\begin{equation}
m\cong\sqrt{8\pi}\left(\frac{R_{f}}{L}\right)^{3/2}\label{eq:m-large_L-result}
\end{equation}
(the so-called ``fluctuational'' or finite-size magnetization).
One can build a good interpolation formula in the whole range of $R_{f}$
that reads
\begin{equation}
m=\frac{1}{\sqrt{1/m_{1}^{2}+1/m_{2}^{2}}},\label{eq:m_disordered_interpolation}
\end{equation}
$m_{1}$ and $m_{2}$ being the two limiting expressions above, Eqs.
(\ref{eq:m-small_L-result}) and (\ref{eq:m-large_L-result}).

One can see that for the $3d$ $xy$ model in Fig. \ref{Fig-m_vs_HR_L=512}
the computed magnetization is substantially larger than the finite-size
magnetization given by Eq. (\ref{eq:m_disordered_interpolation})
in the intermediate range of $H_{R}$. For small $H_{R}$ the system
is short-range-ordered, $m\cong1$. For large $H_{R}$ spins are forced
to align with the random field and the system is completely disordered.
The quasi-plateau of $m(H_{R})$ followed by the shoulder at $H_{R}/J\simeq2.5$
is due to the effect of vortex loops. As further decrease of $m$
requires creating vortex loops that cost energy, the system remains
ordered. \cite{PGC-PRL2014} This holds until the RF overpowers the
exchange.

An interesting fact is the reversibility of the $m(H_{R})$ curve
(in a limited range of $H_{R}$) that contrasts the hysteresis of
$m(H)$ characterized by Barkhausen jumps. As $H_{R}$ increases the
spins turn more and more in the direction of the random field without
overcoming any barriers. This suggests that the obtained state is
generically related to the collinear ground state of a pure ferromagnet.
Reaching $H_{R}/J\simeq2$ causes creation of vortex loops, at which
point the reversibility of the $m(H_{R})$ curve breaks down. Thus
the F-states obtained by the adiabatic increasing of $H_{R}$ are
likely to have the lowest energy among all the F-states.

The results for the $3d$ Heisenberg model ($n=3)$ shown in Fig.
\ref{Fig-m_vs_HR_n=3_L=512_pbc}
reveal a much smaller magnetization of the adiabatic F-state than
for the $xy$ model ($n=2$). Even for the system size as large as
$L=512$, the computed magnetization is not much greater than the
finite-size magnetization. This is because the F-state in the Heisenberg
model is topologically protected by hedgehogs. The latter have much
smaller energy than vortex loops of the $xy$ model. Still, the nearly
constant slope of $m(H_{R})$ in the region $1\lesssim H_{R}/J\lesssim3$
indicates the existence of the ordered state that is slowly destroyed
by emerging singularities -- hedgehogs. The fraction of hedgehogs
$f_{S}$ is shown on the right axis of Fig. \ref{Fig-m_vs_HR_n=3_L=512_pbc}.

\section{Numerical implementation of the Imry-Ma argument}

\label{IM}

Imry and Ma (IM) argument \cite{Imry-Ma-PRL1975} assumes that in
RF models the spins are directed along the random field averaged over
correlated volumes of linear size $R$, found self-consistently by
minimizing the energy due to the random field and exchange. The energy
per spin is estimated as
\begin{equation}
E-E_{0}\sim-sh\left(\frac{a}{R}\right)^{d/2}+s^{2}J\left(\frac{a}{R}\right)^{2},\label{eq:Imry-Ma}
\end{equation}
where $E_{0}$ is the energy per spin of a collinear state. The first
term in this formula is the statistical average of the RF energy in
the IM domain of size $R$ while the second term is the exchange energy
from smooth rotation of the magnetization between adjacent domains.
Minimization of the energy with respect to $R$ yields $R=R_{f}$,
where
\begin{equation}
R_{f}\sim a\left(\frac{sJ}{h}\right)^{2/(4-d)}.\label{eq:Rf-IM-d}
\end{equation}
For $d=3$, up to the dependence on $n$, it coincides with Eq. (\ref{Rf-n}).
The resulting energy of the IM state is
\begin{equation}
E-E_{0}\sim-s^{2}J\left(\frac{h}{sJ}\right)^{4/(4-d)}\label{eq:E-IM-d}
\end{equation}
that yields $E-E_{0}\sim-h^{4}/J^{3}$ in a $3d$ field model.

It can be shown\cite{PGC-PRL2014} that the IM state of the $3d$
$xy$ model inevitably contains vortex loops that cost energy. The
resulting estimation for the exchange becomes\cite{GCP-PRB13}
\begin{equation}
E_{\mathrm{ex}-V}\sim s^{2}J\left(\frac{a}{R}\right)^{2}\ln\left(\frac{R}{a}\right)\label{eq:EexIM_vortex_loops}
\end{equation}
that replaces the second term in Eq. (\ref{eq:Imry-Ma}). Minimization
of the total energy then yields $R=R_{f}$, where
\begin{equation}
R_{f}\sim a\left(\frac{Js}{h}\right)^{2}\ln^{2}\left(\frac{Js}{h}\right)\label{eq:Rf_IM_vortex_loops}
\end{equation}
c.f. Eq. (\ref{eq:Rf-IM-d}). The resulting energy of the IM state
then becomes\cite{GCP-PRB13}
\begin{equation}
E-E_{0}\sim\frac{h^{4}}{s^{2}J^{3}}\frac{1}{\ln^{3}\left(sJ/h\right)}.\label{eq:EIM_vortex_loops}
\end{equation}
In the case of a weak RF the large logarithm in the denominator significantly
decreases the energy gain due to the adjustment of spins to the averaged
RF. It should be stressed that the formula above is only an estimation
and there can be unaccounted large numerical factors, in particular,
in the argument of the logarithm.

In the lattice model the main contribution to the adjustment energy
arizes at the atomic scale and is given by $E-E_{0}\sim-h^{2}/J$
in all dimensions. The IM energy represents small correction to that
energy due to the large-scale rotation of the magnetization on a large
distance $R_{f}$. Finite value of $R_{f}$ for any $d<4$ supports
the IM picture of a disordered state. However, it does not prove rigorously
that the ground state of the system has $m=0$. To prove that one
has to compare the energy of the $m=0$ state, that may contain topological
defects, with the energy of defect-free F-states with $m\neq0$.

Mathematical implementation of the averaged RF is
\begin{equation}
\mathbf{\bar{h}}_{i}=\sum_{j}K_{ij}\mathbf{h}_{j},\label{eq:RF-moving_average_simmation}
\end{equation}
or, within the continuous approximation,
\begin{equation}
\mathbf{\bar{h}}(\mathbf{r})=\intop d^{d}\mathbf{r}'K(r')\mathbf{h}(\mathbf{r}'+\mathbf{r}),\label{eq:RF-moving_average-K}
\end{equation}
where $K$ an averaging kernel. Spins follow the direction of $\mathbf{\bar{h}}$
and have to be normalized,
\begin{equation}
\mathbf{s}(\mathbf{r})=s\frac{\mathbf{\bar{h}}(\mathbf{r})}{\left|\mathbf{\bar{h}}(\mathbf{r})\right|}.\label{eq:s_IM-aligned}
\end{equation}
The choice of the averaging kernel introduces uncertainty into the
IM construction. Possible choices for $K(r)$ are rigid sphere, rigid
cube, Gaussian, exponential, etc.

For the $\delta$-correlated random field, Eqs. (\ref{eq:h-corr-ij})
or (\ref{eq:h-corr_continuous}), the correlator of the averaged RF
is given by
\begin{equation}
\left\langle \bar{h}_{\alpha}(\mathbf{r})\bar{h}_{\beta}(\mathbf{r}')\right\rangle =\frac{h^{2}}{n}\delta_{\alpha\beta}\Gamma(|\mathbf{r}-\mathbf{r}'|),\label{eq:hbar_CF}
\end{equation}
where
\begin{equation}
\Gamma(r)=\frac{1}{a^{d}}\intop d^{d}\mathbf{r}'K(r')K\left(\left|\mathbf{r}-\mathbf{r}'\right|\right).\label{eq:Gamma_via_K}
\end{equation}
Thus the averaged RF is a correlated RF. We will require $\Gamma(0)=1$
that yields the condition
\begin{equation}
1=\frac{1}{a^{d}}\intop d^{d}\mathbf{r}K^{2}(r).\label{eq:Norm}
\end{equation}
Choosing different $K(r)$, one can obtain different $\Gamma(r)$.
Considering, instead of Eq. (\ref{eq:s_IM-aligned}), spin vectors
\begin{equation}
\mathbf{s}(\mathbf{r})=s\frac{\mathbf{\bar{h}}(\mathbf{r})}{h}\label{eq:spins_norm_on_average}
\end{equation}
that are normalized on average amounts to the mean spherical model
(see Sec. \ref{analytical}). The correlation function of these spins
is exactly $\Gamma(r)$. The actual spins, however, have to be normalized
at each site by Eq. (\ref{eq:s_IM-aligned}). The denominator in Eq.
(\ref{eq:s_IM-aligned}) introduces singularities in the spin field
where $\left|\mathbf{\bar{h}}(\mathbf{r})\right|=0$. This happens
forcibly, if the number of spin components $n$ is small enough, $n\leq d$,
that includes practical cases.\cite{PGC-PRL2014} For $n>d$ there
is no topological reason for $\left|\mathbf{\bar{h}}(\mathbf{r})\right|=0$,
and one can expect that the difference between the spin fields defined
by Eqs. (\ref{eq:s_IM-aligned}) and (\ref{eq:spins_norm_on_average})
is small.

Let us consider particular averaging kernels. The rigid-cube kernel
satisfying Eq. (\ref{eq:Norm}) is given by
\begin{equation}
K(\mathbf{r})=\begin{cases}
\left[a/\left(2R\right)\right]^{d/2}, & \left|x\right|,\left|y\right|,\left|z\right|\leq R\\
0 & \left|x\right|,\left|y\right|,\left|z\right|>R.
\end{cases}\label{eq:K_rigid_cut-off}
\end{equation}
The correlation function $\Gamma$ is defined by the overlap area
of two shifted cubes that leads to the small-distance behavior
\begin{equation}
1-\Gamma(r)\sim r/R.\label{eq:Gamma_rigid_cut-off}
\end{equation}
Disappearance of the overlap leads to the vanishing of $\Gamma$ at
finite distances, an undesirable property.

The Gaussian averaging kernel in $3d$ is given by
\begin{equation}
K(r)=\left(\frac{2}{\sqrt{\pi}}\frac{a}{R}\right)^{3/2}e^{-2\left(r/R\right)^{2}}.\label{eq:K_Gaussian}
\end{equation}
The corresponding RF correlation function is
\begin{equation}
\Gamma(r)=e^{-\left(r/R\right)^{2}}.\label{eq:CF-Gaussian}
\end{equation}
One can see that this result disagrees with the spin-spin correlation
function that follows from the Green-function method, as well as
with Eq. (\ref{expCF}) for the mean spherical model.

The exponential averaging kernel in $3d$ is given by
\begin{equation}
K(r)=\frac{1}{\sqrt{\pi}}\left(\frac{a}{R}\right)^{3/2}e^{-r/R}.\label{eq:K_exp}
\end{equation}
The corresponding RF correlation function is non-exponential. It goes
quadratically at small distances,
\begin{equation}
\Gamma(r)\cong1-\frac{1}{6}\left(\frac{r}{R}\right)^{2},\qquad r\ll R\label{eq:Gamma_K_exp_short_r}
\end{equation}
again contradicting and has the long-distance behavior
\begin{equation}
\Gamma(r)\cong\frac{1}{3}\left(\frac{r}{R}\right)^{2}e^{-r/R},\qquad R\ll r,\label{eq:Gamma_K_exp_large_r}
\end{equation}
again, contradicting Eq. (\ref{expCF}).

On the contrary, Yukawa averaging kernel in $3d$
\begin{equation}
K(r)=\frac{a^{3/2}}{\sqrt{2\pi R}}\frac{1}{r}e^{-r/R}\label{eq:K_Yukawa}
\end{equation}
that has the same singularity as the Green function of the mean-spherical
model $(n=\infty$), leads to the desired exponential correlation
function of the correlated RF
\begin{equation}
\Gamma(r)\cong e^{-r/R}\label{eq:Gamma-LD}
\end{equation}
and thus of the spin-spin correlation function for spins normalized
on average. This is not surprising because for the Yukawa kernel Eqs.
(\ref{eq:RF-moving_average-K}) and (\ref{eq:spins_norm_on_average})
are equivalent to Eq. (\ref{eq:S_via_h_mean_spherical}) that results
in Eq. (\ref{expCF}).

\begin{figure}
\includegraphics[width=8cm]{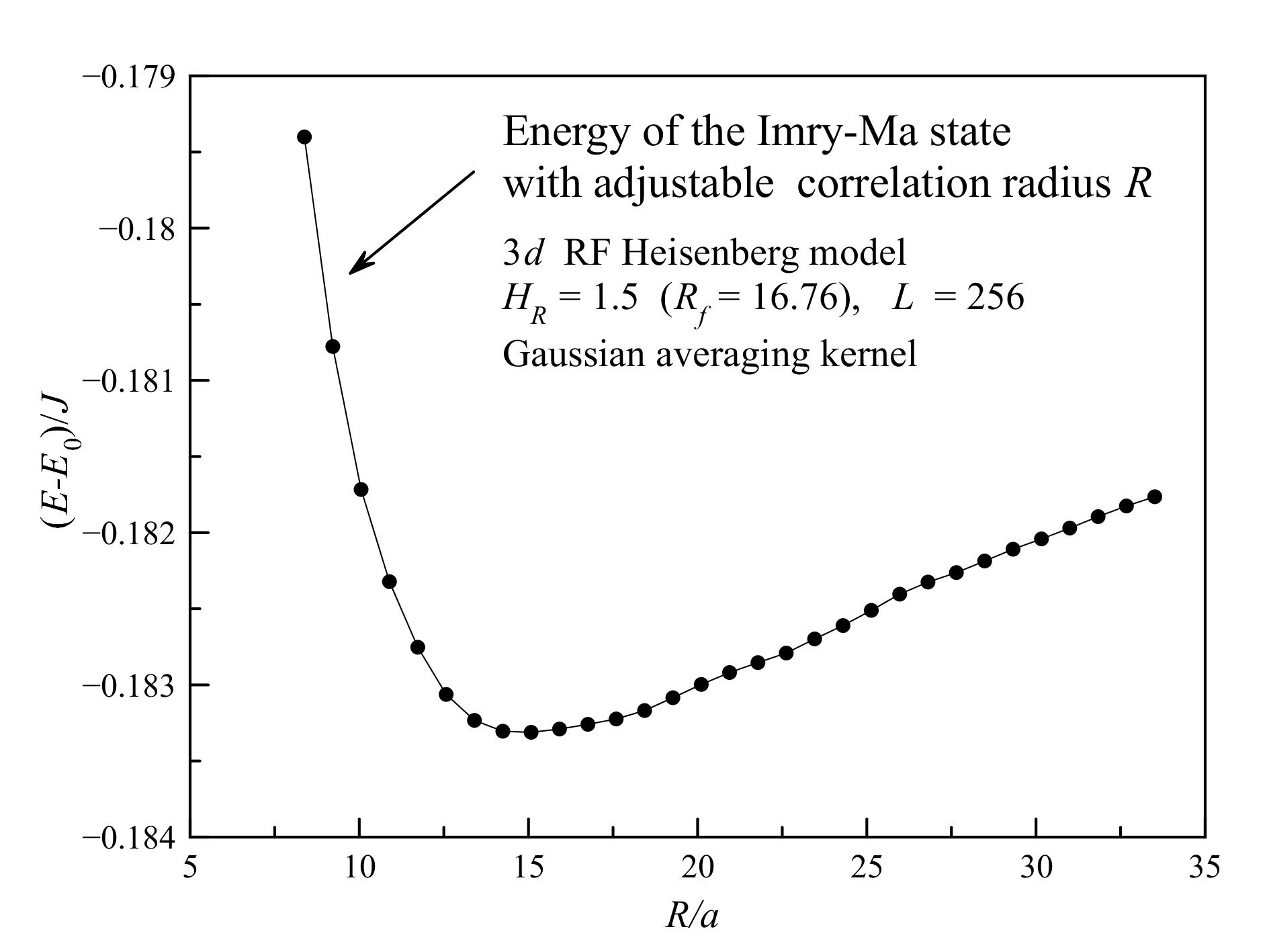}

\caption{Energy of the constructed IM state in the $3d$ Heisenberg model vs
the averaging range $R$. Gaussian averaging kernel is used and the
short-range energy of Eq. (\ref{E-SR}) is added. }

\label{Fig-EIM_vs_R_Nalp=3_Nx=Ny=Nz=256_HR=1.5_pbc}
\end{figure}

\begin{figure}
\includegraphics[width=8cm]{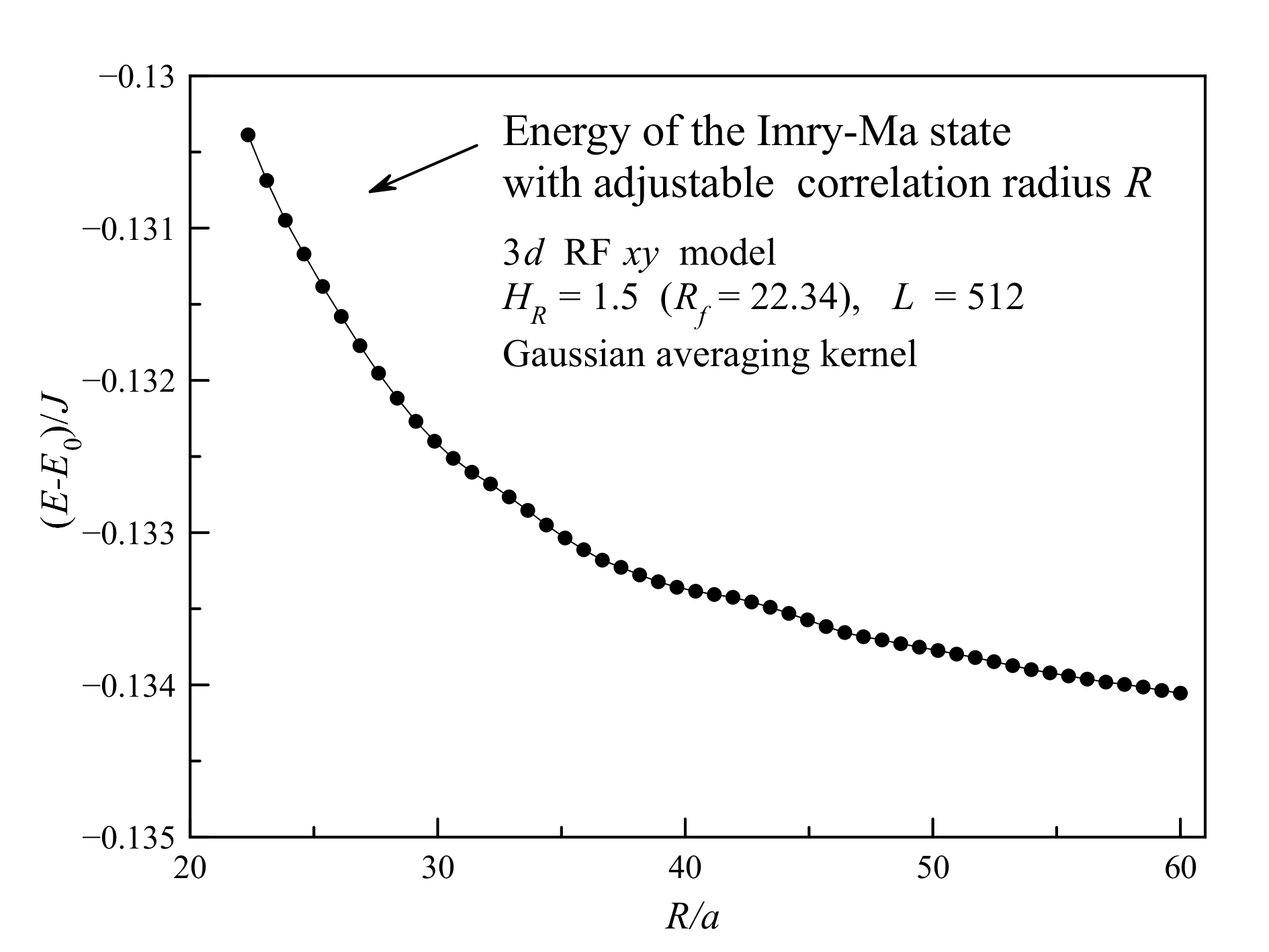}

\caption{Energy of the constructed IM state in the $3d$ $xy$ model vs the
averaging range $R$. Gaussian averaging kernel is used and the short-range
energy of Eq. (\ref{E-SR}) is added. }

\label{Fig-EIM_vs_R_Nalp=2_Nx=Ny=Nz=512_HR=1.5_pbc}
\end{figure}

Let us consider the energies of the constructed IM states, using the
simplified form of the spin field given by Eq. (\ref{eq:spins_norm_on_average}).
Then $\Gamma(r)$ above is the spin-spin correlation function. The
exchange energy is defined by the nearest-neighbor correlation function,
thus
\begin{equation}
E_{\mathrm{ex}}-E_{0}=s^{2}dJ\left[1-\Gamma(a)\right],\label{eq:Eex_general}
\end{equation}
where $E_{0}=dJ$ is the ground-state energy of a pure magnet. This
method of computing the exchange energy used in Ref. \cite{GCP-PRB13}
ignores lattice-discreteness effects at small distances. However,
it can be shown that the error is very small. One can see that $E_{\mathrm{ex}}-E_{0}\propto1/R$
for the rigid cut-off and Yukawa averaging kernels and $E_{\mathrm{ex}}-E_{0}\propto1/R^{2}$
for the Gaussian and exponential averaging kernels. As the second
case is in accord with the IM argument, this type of kernels will
be called regular.

The random-field energy per spin is given by $E_{RF}=-\left\langle \mathbf{s}_{i}\mathbf{\cdot h}_{i}\right\rangle $.
Using Eq. (\ref{eq:spins_norm_on_average}), one obtains
\begin{eqnarray}
E_{RF} & = & -\frac{s}{h}\left\langle \mathbf{\bar{h}}_{i}\mathbf{\cdot h}_{i}\right\rangle =-\frac{s}{h}\left\langle \sum_{j}K_{ij}\mathbf{h}_{j}\mathbf{\cdot h}_{i}\right\rangle \nonumber \\
 & = & -shK_{ii}=-shK(0).\label{eq:ERF_general}
\end{eqnarray}
For regular kernels as well as for the rigid cut-off kernel in $3d$,
using Eqs. (\ref{eq:K_Gaussian}) and (\ref{eq:K_exp}), one obtains
$E_{RF}\propto-1/R^{3/2}$, as in the IM argument. In fact, there
is an agreement of both exchange and random-field energies with their
IM forms for any dimension $d$, if regular averaging kernels are
used. Minimizing the total energy with respect to $R$, one obtains
the value of $R_{f}$ above and the minimal energy $E-E_{0}\propto-h^{4}/J^{3}$
in $3d$, according to Eq. (\ref{eq:E-IM-d}). Thus, regular averaging
kernels fully reproduce the original IM argument \cite{Imry-Ma-PRL1975}
and provide its implementation with the results differing only by
a numerical factor.

However, using regular kernels one misses the main contribution to
the energy due to the adjustment of spins at the atomic scale, $E-E_{0}\propto-h^{2}/J$.\cite{GCP-PRB13}
Another drawback of these kernels is the wrong spin-spin correlation
function in $3d$.

For the Yukawa averaging kernel, Eq. (\ref{eq:ERF_general}) is singular
within the continuous approximation. However, the lattice Green function
with coinciding indices is finite. The regularization the can be done
by the replacement $r\rightarrow a$ in the denominator of Eq. (\ref{eq:K_Yukawa}).
The corresponding modification yields
\begin{equation}
K_{ii}=\eta\sqrt{\frac{a}{2\pi R}},\label{eq:K_Yukawa_ii}
\end{equation}
where $\eta$ is a constant. The total energy for the Yukawa averaging
kernel following from Eqs. (\ref{eq:Gamma-LD}), (\ref{eq:Eex_general}),
(\ref{eq:ERF_general}) and the formula above in $3d$ is given by
\begin{equation}
E-E_{0}=3s^{2}J\frac{a}{R}-sh\eta\sqrt{\frac{a}{2\pi R}}.\label{eq:E-Yukawa}
\end{equation}
Minimizing this energy over $R$ yields
\begin{equation}
\frac{R_{f}}{a}=8\pi\left(\frac{3J}{\eta h}\right)^{2}
\end{equation}
and
\begin{equation}
E_{\mathrm{ex}}-E_{0}=\frac{\eta^{2}h^{2}}{24\pi J},\qquad E_{RF}=-2\left(E_{\mathrm{ex}}-E_{0}\right).
\end{equation}
Choosing $\eta^{2}=9\left(1-1/n\right)$, one recovers $R_{f}$ of
Eq. (\ref{eq:Rf-IM-d}) and the total adjustment energy of Eq. (\ref{E-SR}).
Summarizing, using the Yukawa averaging kernel in the IM construction
provides correct forms of the spin correlation function, including
the magnetic correlation length $R_{f}$, as well as the leading contribution
to the energy $h^{2}/J$. In the limit $n\rightarrow\infty$ the results
above become exact and coincide with those of the mean spherical model.

The approach proposed above allows to numerically minimize the energy of the system on the magnetic
correlation range, i.e., the averaging range $R$. As the spirit of the original Imry-Ma argument requires a non-singular averaging kernel,
we will use the Gaussian kernel that reproduces both terms of the basic equation (\ref{eq:Imry-Ma}).
For the $3d$ Heisenberg model in Fig. \ref{Fig-EIM_vs_R_Nalp=3_Nx=Ny=Nz=256_HR=1.5_pbc},
there is clearly the energy minimum at $R/a\simeq15$ that is close
to $R_{f}/a=16.76$. One can see that singularities of the spin field,
hedgehogs, do not disturb much the IM argument. On the contrary, for
the $3d$ $xy$ model, we were unable to find the energy minimum for
$H_{R}/J=1.5$ up to the rather large values of $R$, as can be seen
in Fig. \ref{Fig-EIM_vs_R_Nalp=2_Nx=Ny=Nz=512_HR=1.5_pbc}.
This striking effect can be attributed to vortex loops that inevitably
arize in the completely disordered $3d$ $xy$ model\cite{PGC-PRL2014}
and increase the exchange energy by a large factor $\ln\left(R/a\right)$,
Eq. (\ref{eq:EexIM_vortex_loops}), making it decrease slower with
$R$. On the top of it, one has to suggest a large numerical factor
in the argument of the logarithm that is absent in the simple estimation
above. The bottom line is that numerical implementation of the Imry-Ma construction with "traditional" non-singular
averaging kernels does not work for the $3d$ $xy$ model. This is not a surprise given that non-singular kernels result in a wrong
spin correlation function.

\begin{figure}
\includegraphics[width=8cm]{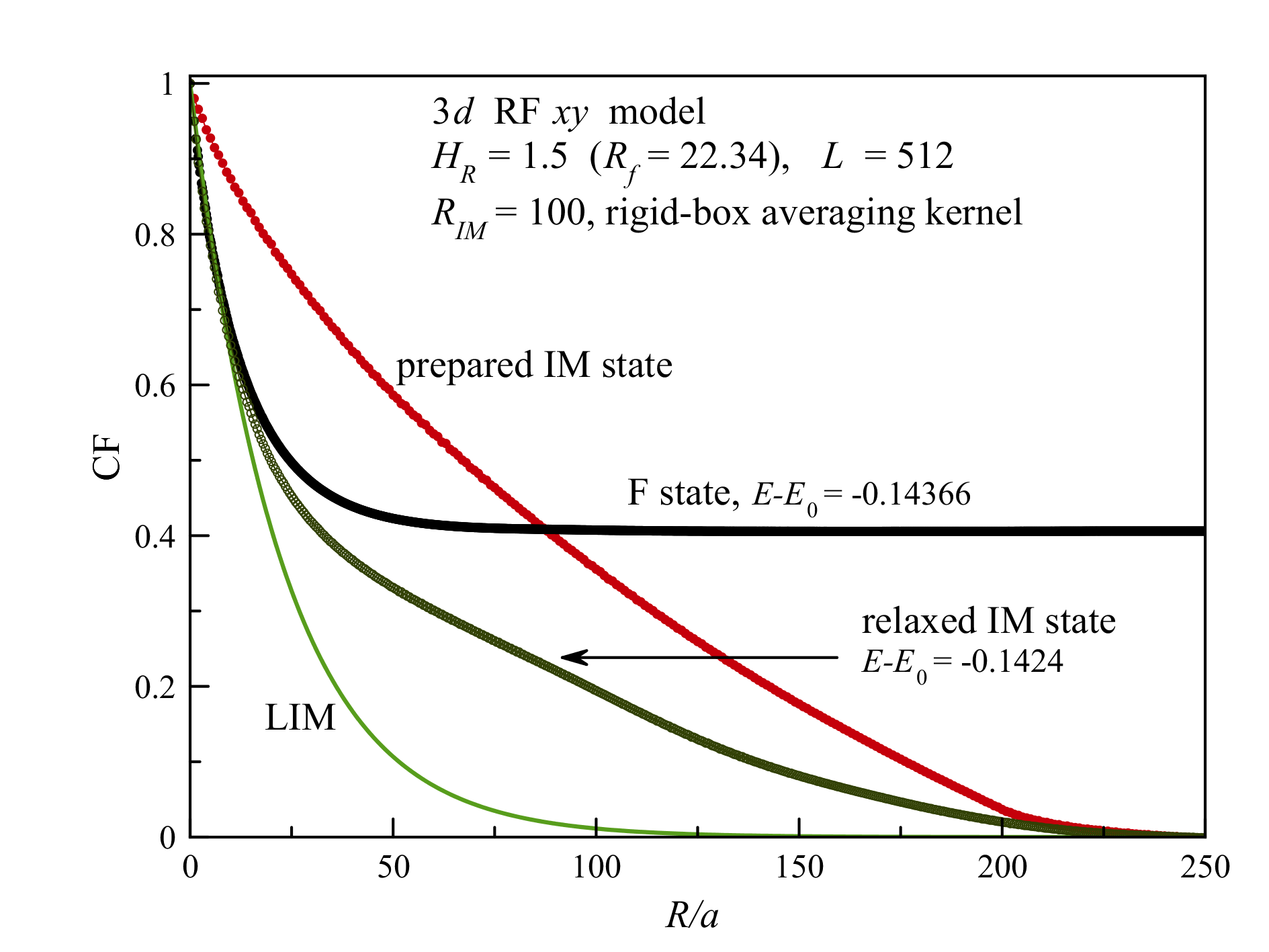}

\caption{Spin-spin correlation function of the constructed IM state with the
rigid-cube cut-off and of the corresponding relaxed state. The spin-spin
correlation function of the F-state and the theoretical LIM curve
are shown for comparison.}
\label{Fig-CF-IM_and_relax_vs_n_Nalp=2_Nx=Ny=Nz=512_HR=1.5_pbc_R=100_rigid_box}

\end{figure}

\begin{figure}
\includegraphics[width=8cm]{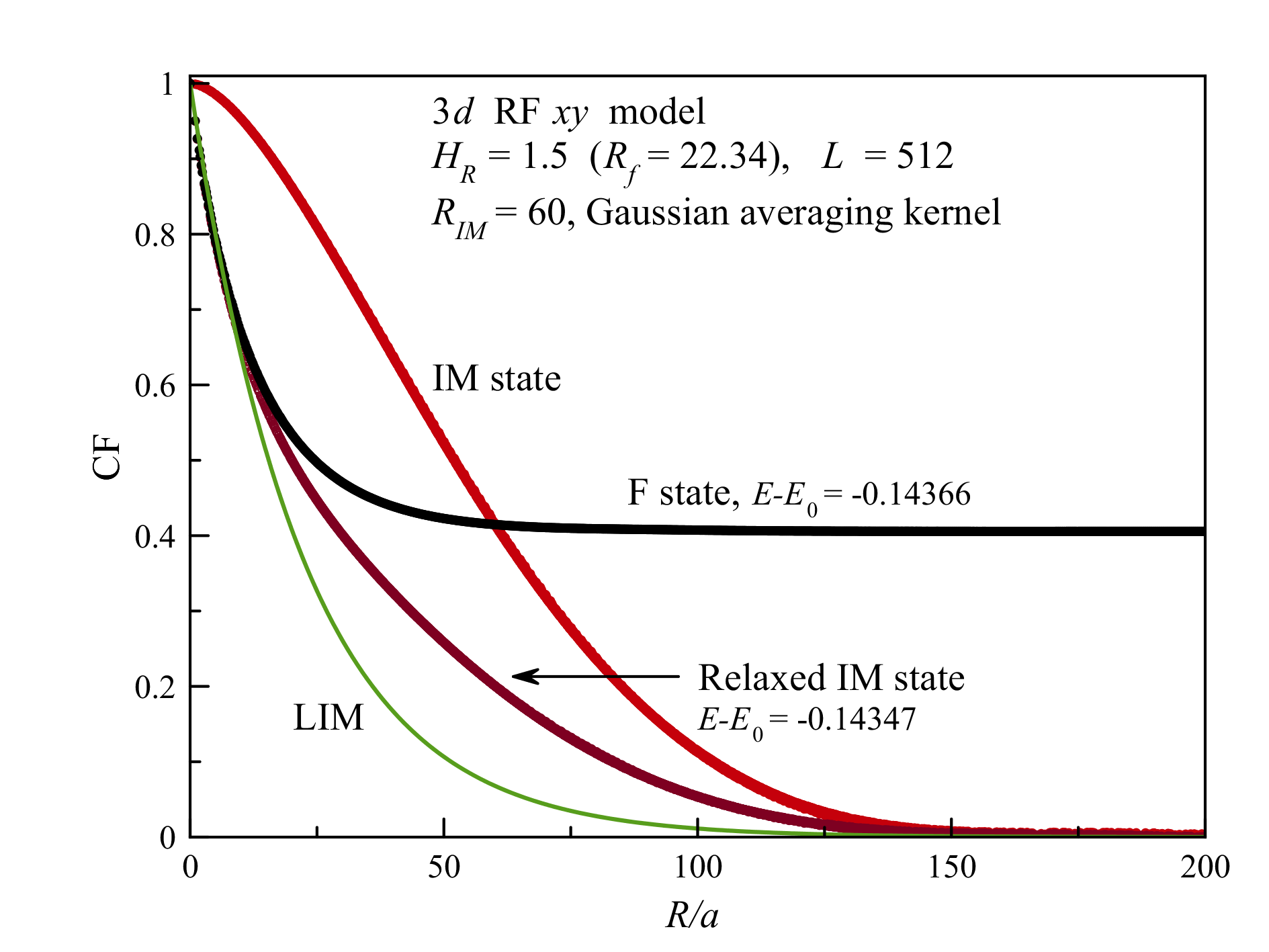}

\caption{Spin-spin correlation function of the constructed IM state with the
Gaussian averaging kernel and of the corresponding relaxed state.
The spin-spin correlation function of the F-state and the theoretical
LIM curve are shown for comparison.}

\label{Fig-CF-IM_and_relax_vs_n_Nalp=2_Nx=DNy=Nz=512_HR=1.5_pbc_R=60}
\end{figure}

\begin{figure}
\includegraphics[width=8cm]{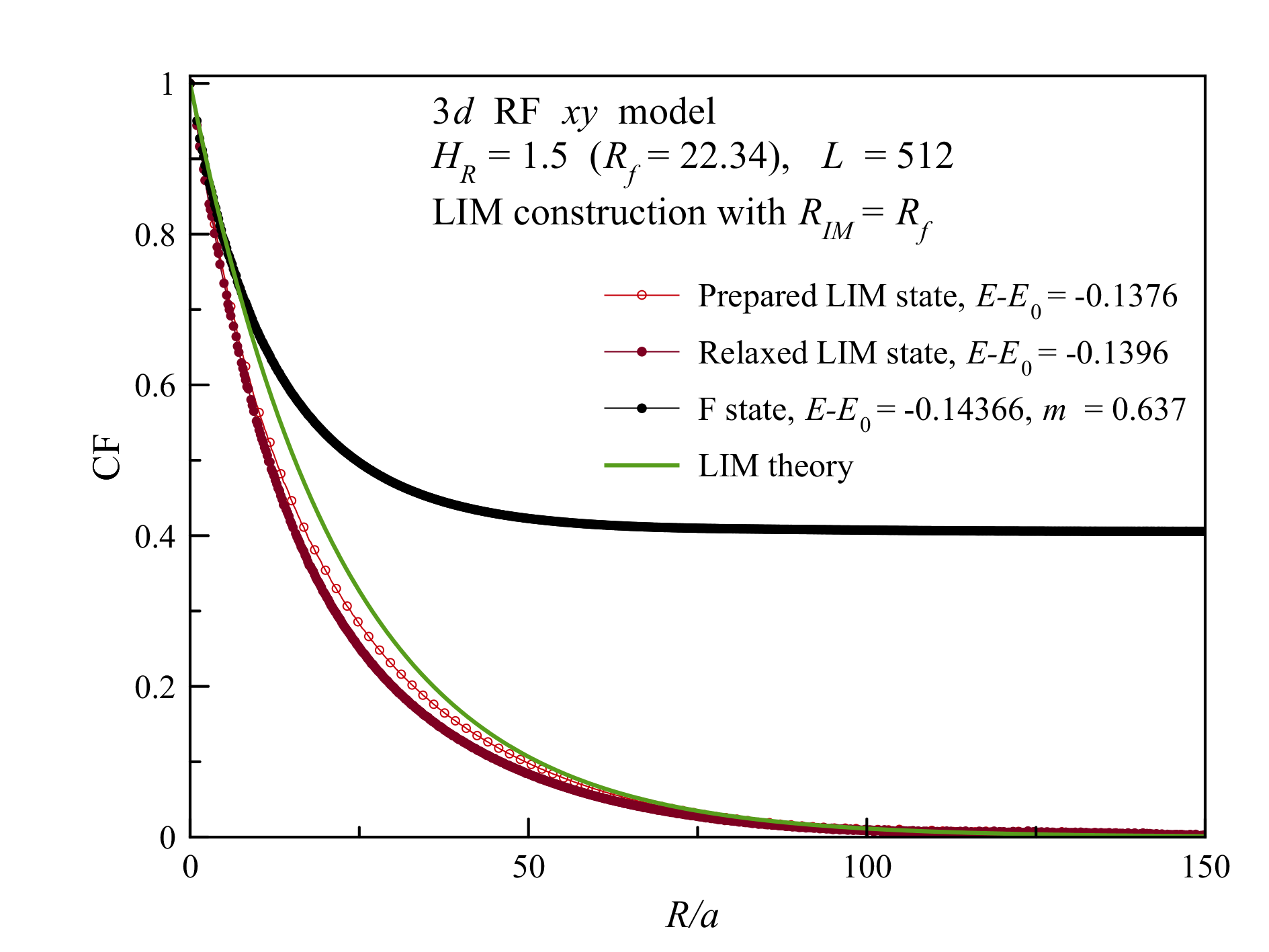}

\caption{Spin-spin correlation function of the constructed IM state with the
Yukawa averaging kernel and of the corresponding relaxed state. The
spin-spin correlation function of the F-state and the exponential
LIM curve are shown for comparison.}

\label{Fig-CF-IM-LIM_and_relax_vs_n_Nalp=2_Nx=Ny=Nz=512_HR=1.5_pbc_R=Rf}
\end{figure}

As non-singular averaging kernels do not describe the short-range adjustment energy $h^2/J$, Imry-Ma states obtained with the
help of them possess a too high energy. Thus they have to be relaxed numerically to reach the actual energy corresponding to a local energy minimum.
Below we will show the results for the rigid-cut-off, Gaussian, and Yukawa kernels.

The spin-spin correlation function for the rigid-cut-off state with $R=R_{IM}=100$
is shown in Fig. \ref{Fig-CF-IM_and_relax_vs_n_Nalp=2_Nx=Ny=Nz=512_HR=1.5_pbc_R=100_rigid_box}.
It has a characteristic shape with a nearly-constant slope ending
at the distance $2R$. This shape is very different from the IM exponential
correlation function that follows from the Green function method.
Starting from this constructed IM state, numerical minimization of
the energy with $H_{R}/J=1.5$ was performed. The resulting spin-spin
correlation function also shown in the figure coincides with the IM
curve and with the correlation function of the F-state at the distances
$r\lesssim R_{f}$, where $R_{f}/a=22.34$. At larger distances this
correlation function is decreasing slowly, similarly to the constructed
IM correlation function, and it turns to zero at the same distance.
This result implies that at distances $r\lesssim R_{f}$ the system
relaxes to minimize its energy, whereas at larger distances the system
is pinned by the RF and its state depends on the initial state.

Similar
results were obtained with the Gaussian averaging kernel, see Fig.
\ref{Fig-CF-IM_and_relax_vs_n_Nalp=2_Nx=DNy=Nz=512_HR=1.5_pbc_R=60}.
Energies of the relaxed IM states in both figures above are slightly
greater than the energy of the F-state. This means that long-distance
correlations can be changed at a very small energy cost, similarly
to creating a domain wall in a big system.

The results for the IM construction with the Yukawa averaging kernel
for the $3d$ $xy$ model are shown in Fig. \ref{Fig-CF-IM-LIM_and_relax_vs_n_Nalp=2_Nx=Ny=Nz=512_HR=1.5_pbc_R=Rf}.
Here also $H_{R}/J=1.5$ was used, and the constructed state was created
with the actual spin correlation range $R_{f}/a=22.34$. Since spins
are normalized by Eq. (\ref{eq:s_IM-aligned}) this creates singularities,
the spin-spin correlation function goes slightly below the LIM curve,
as can be seen in Fig. \ref{Fig-CF-IM-LIM_and_relax_vs_n_Nalp=2_Nx=Ny=Nz=512_HR=1.5_pbc_R=Rf}.
Subsequent relaxation makes the correlation function go slightly lower
than the IM curve. The energies of both, the constructed IM state
and the corresponding relaxed state, are close to that of the F-state
but still higher. This is because the constructed IM state and the
relaxed state possess vortex loops that cost energy.

One can conclude that the singular Yukawa
averaging kernel provides a much better approach to the structure of the disordered magnetic states than
non-singular averaging kernels in the spirit of the original Imry-Ma argument.

Energies of the relaxed states obtained from the constructed IM states
with a large correlation range, see Figs. \ref{Fig-CF-IM_and_relax_vs_n_Nalp=2_Nx=Ny=Nz=512_HR=1.5_pbc_R=100_rigid_box}
and \ref{Fig-CF-IM_and_relax_vs_n_Nalp=2_Nx=DNy=Nz=512_HR=1.5_pbc_R=60},
are slightly above the energy of the F-state. These states have a
two-scale structure defined by $R_{f}$ of Eq. (\ref{eq:Rf-IM-d})
and the initial large correlation range. As the energy of pinned state
at the scale $R\gtrsim R_{f}$ in the computations above is not minimized,
one can imagine that one could find a state with the energy that is
lower than that of the F-state. This presumably can be done by slowly
rotating the F-state on a characteristic scale $R'\gg R$. At first
glance it appears obvious that this would win an additional Zeeman
energy $-\mathbf{m}\cdot\left\langle \mathbf{h}\right\rangle $, where
$\left\langle \mathbf{h}\right\rangle $ is the random field averaged
on the scale $R'$. The obvious candidate for $R'$ in a $3d$ $xy$
model is the IM length with account of one vortex loop per IM domain,
Eq. (\ref{eq:Rf_IM_vortex_loops}). Such a ground state would have
zero magnetization. It would have two scales: $R_{f}$ characterizing
the partial disordering of the magnetization inside F-state domains
of size$R'\gg R$ that are oriented randomly with respect to each
other, with a smooth rotation of $\mathbf{m}$ between the domains.
Notice that in the absence of a proof, one also cannot exclude the
possibility that in this two-scale ground state the magnetization
of the F-state domains rotates randomly with a power-law decay of
correlations at large distances as predicted by the Bragg glass theory.
It should be noted that we do not have an analytical theory of this
two-scale state. Also it is very difficult to obtain numerically becase
of the energy barriers and a very small energy gain.

\section{Microscopic structure of the F-state}

\label{structure}

For arbitrary number of spin components $n$ the correlation range
$R_{f}$ of the RF system perfectly agrees with the value calculated
by the Green-function method, that is itself in a qualitative agreement
with the Imry-Ma argument. For any $n$ the spin-spin correlation
function decreases as $C(r)\cong1-r/R_{f}$ at distances $r\lesssim R_{f}$.
At $n>d+1$ correlations decay exponentially regardless of the initial
condition for the spins. \cite{PGC-PRL2014} However, for $n\leq d$
at $r>R_{f}$ the correlation function of the state obtained by the
relaxation from the fully ordered state has a plateau corresponding
to a non-zero magnetization. The spin field in the F-state does not
have singularities that would arise in the case of $n\leq d$ in the
hypothetical Imry-Ma state in which spins follow the RF averaged inside
the IM domains of linear size $R_{f}$.

\begin{figure}
\includegraphics[width=8cm]{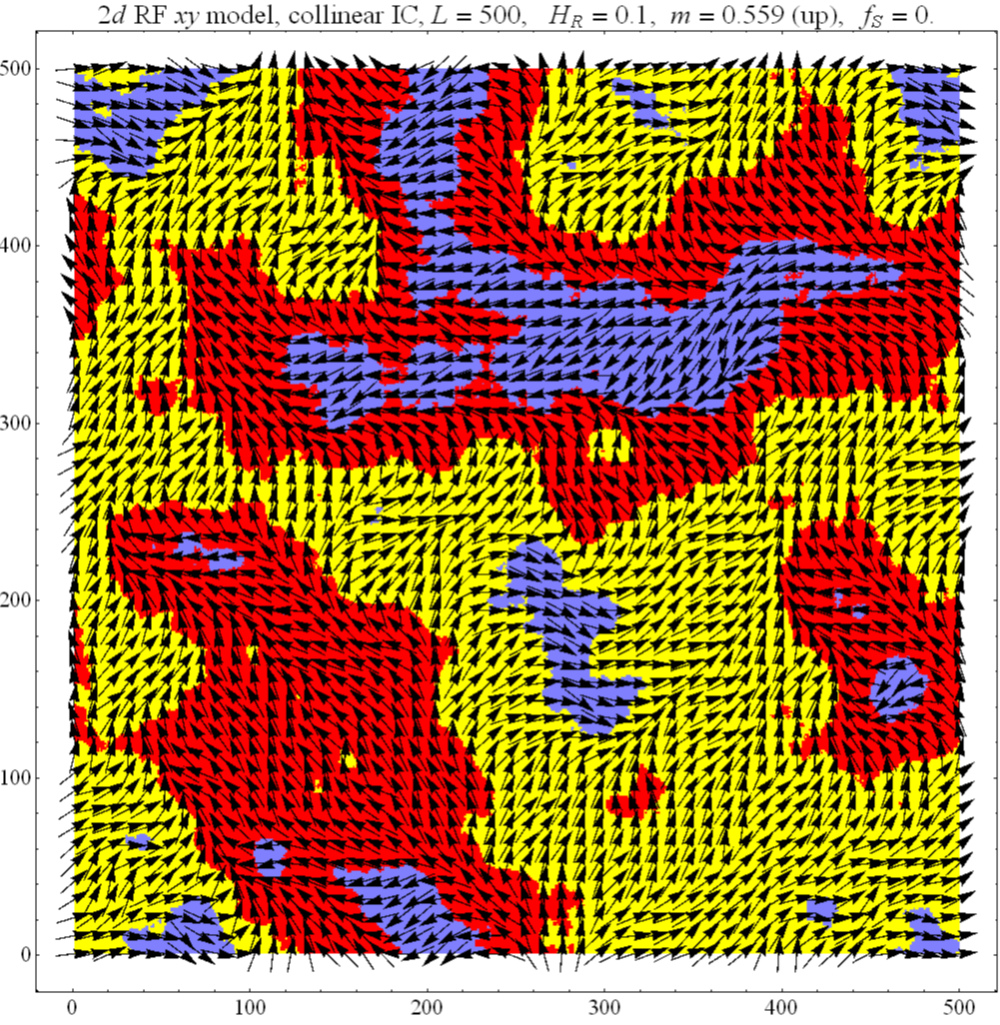}

\caption{Imry-Ma domains in the F-state of the $2d$ $xy$ model at zero applied field. Only a fraction of spins are shown.}

\label{Spins_2d_RF_xy_model_HR=0.1_L=500_coll_IC}
\end{figure}

It is interesting to look at the difference between the F-state and
the IM state by vizualising the domains. For this purpose, color labeling
of the F-states in the $2d$ $xy$ model
obtained, as usual, by relaxation from a collinear state in zero applied field,
has been done as shown in
Fig. \ref{Spins_2d_RF_xy_model_HR=0.1_L=500_coll_IC}.
With the sample's magnetization pointing up (positive $y$ direction),
regions with spins turned to the right are coded yellow and spins
turned to the left are coded red. Spins directed down are labeled
blue overriding yellow and red. The results can be understood as follows.
Under the influence of the random field spins are rotating clockwise
and counterclockwise from the initial direction up, becoming yellow
and red. Rotating more into the down region, they become blue. In
this scenario, blue regions are entirely inside their parent yellow
and red regions and spins are divided into rotated clockwise and counterclockwise.
This spin state is topologically equivalent to the initial collinear
state and can be transformed back to it without creating or annihilating
topological structures.

\begin{figure}
\includegraphics[width=8cm]{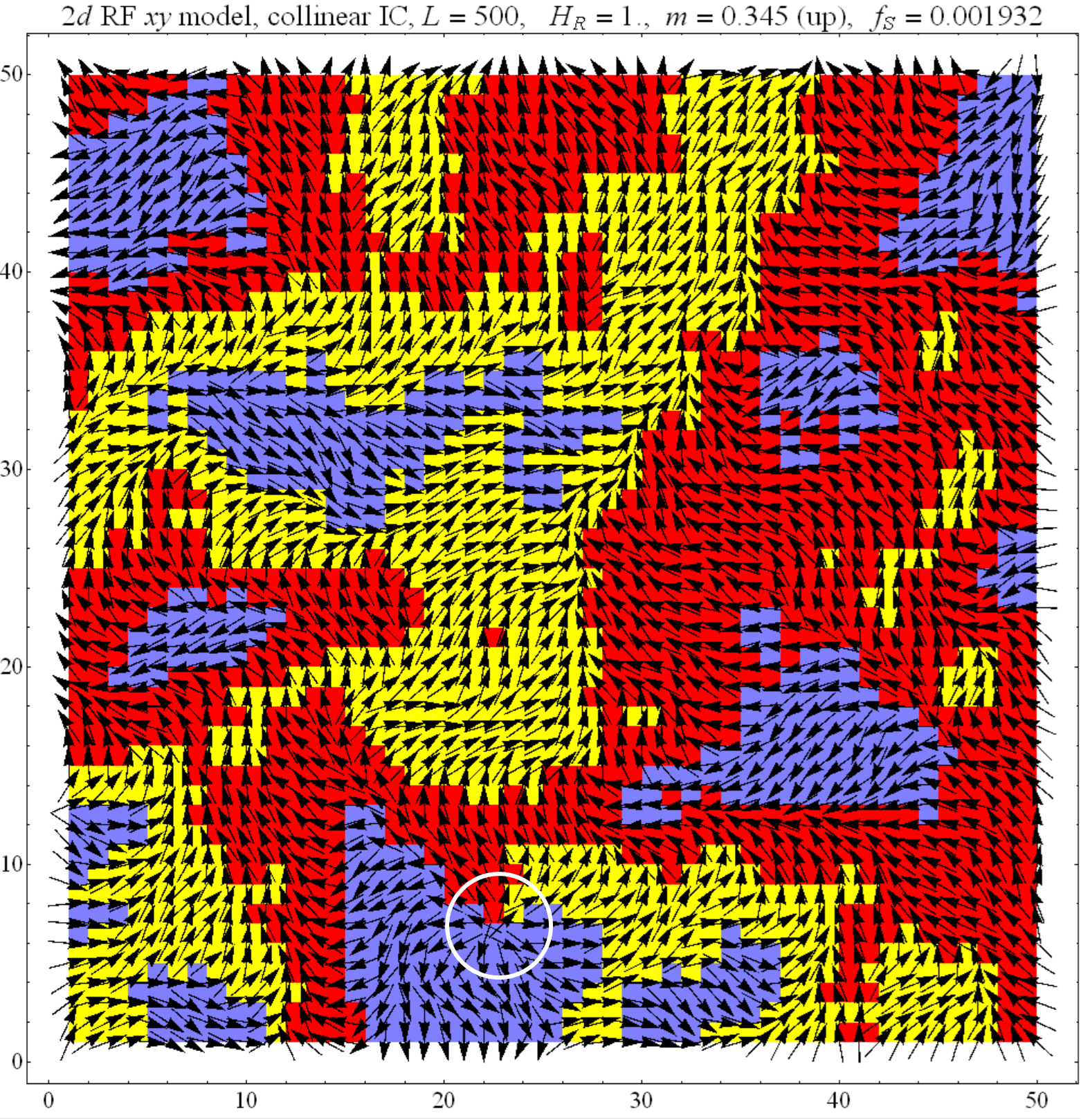}

\caption{Imry-Ma domains in the F-state of the $2d$ $xy$ model at zero applied field, with a vortex
in the lower part inside the white circle. Only a fraction of the lattice is shown.}

\label{Spins_2d_RF_xy_model_HR=1_L=500_coll_IC-fragment}
\end{figure}

Color-coded regions in Fig. \ref{Spins_2d_RF_xy_model_HR=0.1_L=500_coll_IC}
can be interpreted as IM domains. Boundaries between yellow and red
correspond to $s_{x}=0$, whereas the boundaries between yellow and
blue or between red and blue correspond to $s_{y}=0$. As was shown
in Ref.\ \cite{GCP-PRB13} the IM construction inevitably
leads to the formation of vortices or anti-vortices on purely mathematical
grounds. This does not happen when spins rotate clockwise or counterclockwise
to form the F-state. In that state the spins do not follow the averaged
random field precisely and the system disorders only partially, keeping
the memory of the initial collinear state. Complete disordering is
blocked because singularities would increase the system's energy.

If blue regions corresponding to the down spins rotated clockwise
and counterclockwise merge, as can be seen near the bottom of Fig.
\ref{Spins_2d_RF_xy_model_HR=1_L=500_coll_IC-fragment},
there would be an intersection of the yellow-red boundary with the
red-blue and yellow-blue boundaries, that is, an intersection of the
$s_{x}=0$ and $s_{y}=0$ lines. As the result, there is a white-encircled
vortex in Fig. \ref{Spins_2d_RF_xy_model_HR=1_L=500_coll_IC-fragment}.
Singularities, such as vortices, can be created by a strong enough
random field out of a collinear state, as is the case in Fig. \ref{Spins_2d_RF_xy_model_HR=1_L=500_coll_IC-fragment}.
One can see that the state shown in Fig. \ref{Spins_2d_RF_xy_model_HR=1_L=500_coll_IC-fragment}
has lost the memory of the initial collinear state and it cannot be
returned to it without changing topology.

One can suggest the form of the IM argument that does not assume a
complete disordering and is thus suitable for the description of F-states.
\cite{GCP-PRB13} It takes into account the adjustment of spins to
the random field at all length scales. Groups of spins of linear size
$R$ rotate by an adjustment angle $\phi$ (considered as small to
begin with) under the influence of the components of the averaged
random field perpendicular to the initial direction. The corresponding
energy per spin is given by
\begin{equation}
E-E_{0}\sim-sh\left(\frac{a}{R}\right)^{d/2}\phi+s^{2}J\left(\frac{a}{R}\right)^{2}\phi^{2}.\label{eq:Imry-Ma-phi}
\end{equation}
Minimizing this expression with respect to $\phi$, one obtains
\begin{equation}
\phi\sim\left(\frac{R}{R_{f}}\right)^{(4-d)/2},\label{eq:phi-IM}
\end{equation}
where $R_{f}$ is given by Eq. (\ref{eq:Rf-IM-d}). The angular deviation
increases with the distance and becomes large at $R\sim R_{f}.$ The
energy per spin corresponding to spin adjustment at the distance $R$
can be obtained by substituting Eq. (\ref{eq:phi-IM}) into Eq. (\ref{eq:Imry-Ma-phi}).
The result has the form
\begin{equation}
E-E_{0}\sim-\frac{h^{2}}{J}\left(\frac{a}{R}\right)^{d-2}.\label{eq:Energy-IM-d-R}
\end{equation}
One can see that the highest energy gain is provided by spin adjustments
at the atomic scale, $R\sim a$. In this case one obtains
\begin{equation}
E-E_{0}\sim-h^{2}/J.\label{eq:E-atomic}
\end{equation}
Substituting $R_{f}$ of Eq. (\ref{eq:Rf-IM-d}) into Eq. (\ref{eq:Energy-IM-d-R}),
one recovers the IM energy of Eq. (\ref{eq:E-IM-d}). One can see
that the IM energy in the F-state is of the same order of magnitude
as the regular IM energy. Because of the incomplete adjustment of
spins to the random field in the F-state, one can expect that its
IM energy contains a smaller numerical factor than Eq. (\ref{eq:E-IM-d}).
Nevertheless, this energy should be lower than Eq. (\ref{eq:EIM_vortex_loops})
because of the logarithmic factor in the latter.

\section{Memory of the initial condition: Rotational elasticity of the F-state}

\label{rotational}

The magnetization in the F-state can be pointed in any direction because
of the macroscopic isotropy of the problem. This direction practically
coincides with the magnetization direction in the initial collinear
state. Macroscopic isotropy does not mean, however, that the magnetization
in the F-state can be rotated at no energy cost, as it would be in
the isotropic ferromagnet in the absence of the RF. The difference
is that the magnetization in the F-state is pinned by a partial adjustment
to the random field. It has a memory of the initial collinear state.
Rotation of the magnetization of the sample, even by a small angle,
requires readjustment of the local magnetization that is inhibited
by local energy barriers.

\begin{figure}
\includegraphics[width=8cm]{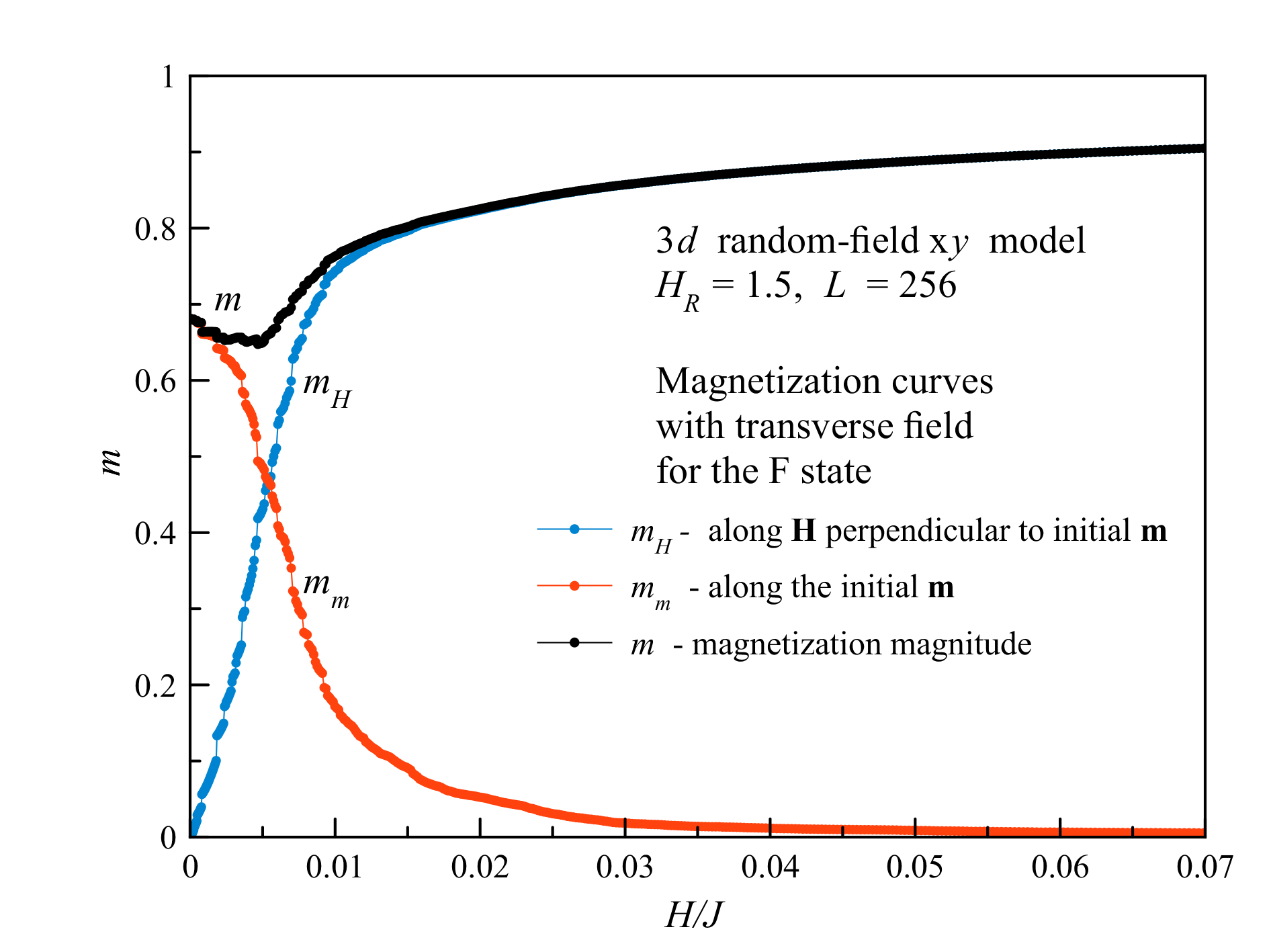}

\caption{Magnetization curves of the F-state in the $3d$ $xy$ model in a
transverse field.}

\label{Fig-mH_mm_m_vs_Htr_for_F_state_Nalp=2_Nx=Ny=Nz=256_HR=1.5}
\end{figure}

An illustration of this memory effect is the dependence of magnetization
components on the magnetic field perpendicular to the initial magnetization
shown in Fig. \ref{Fig-mH_mm_m_vs_Htr_for_F_state_Nalp=2_Nx=Ny=Nz=256_HR=1.5}.
The magnetization component in the direction of the field $m_{H}$
increases from zero with a \textit{finite slope} because of pinning,
whereas the magnetization component along the initial magnetization
$m_{m}$ decreases to zero. After aligning with the field at larger
fields, the magnetization increases. One can see Barkhausen jumps
in the magnetization curves. For smaller or larger values of $H_{R}$
the dependence $m_{H}$ is less linear and it is difficult to scale
the results neatly.

\begin{figure}
\includegraphics[width=8cm]{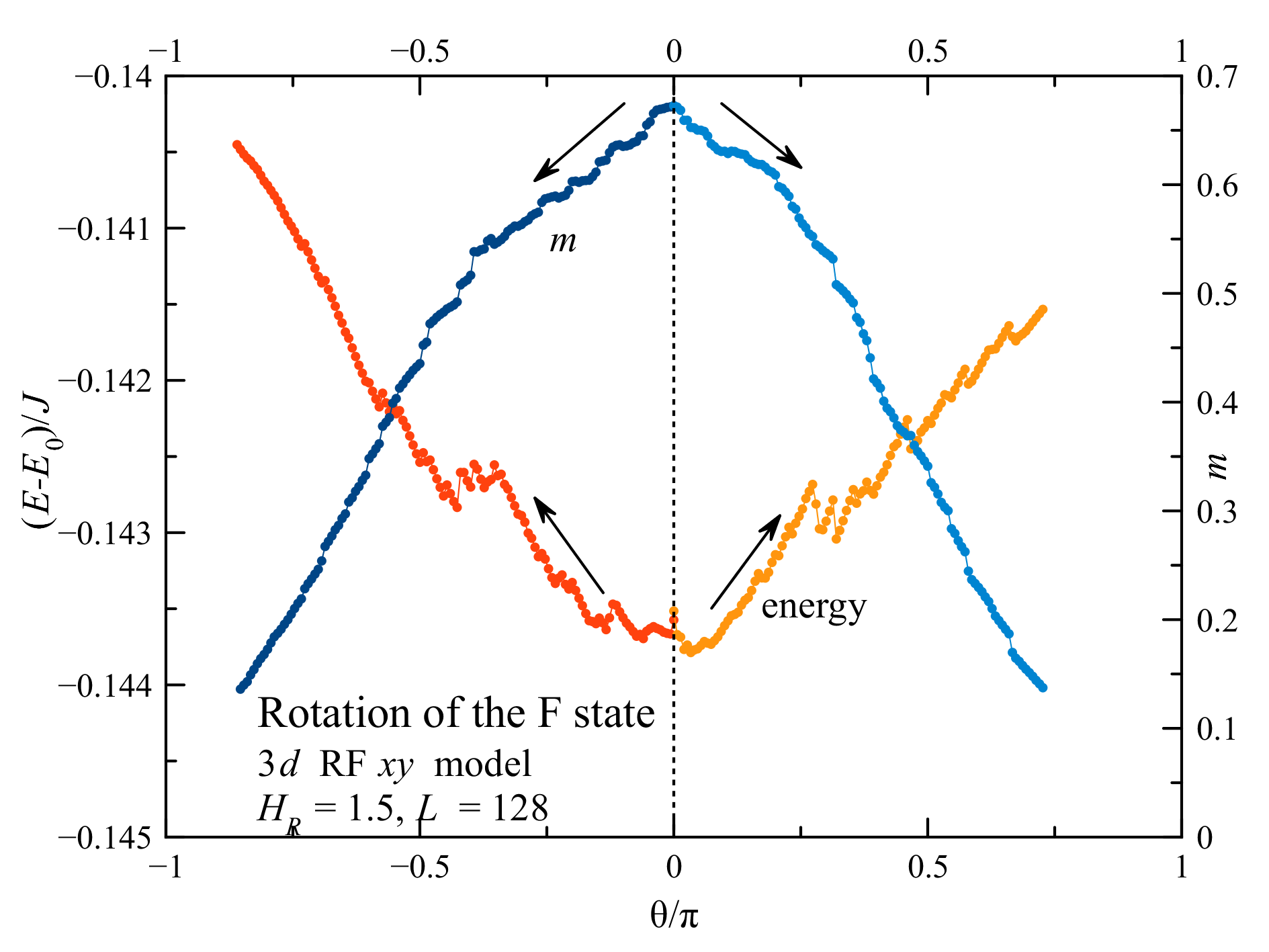}

\caption{Rotation of the F-state.}

\label{Fig-dE_and_m_vs_theta_Nalp=2_L=128_HR=1.5}
\end{figure}

Another, more pure, numerical experiment on the rotation of the magnetization
in the F-state can be performed as follows. An F-state is prepared
by the relaxation from a collinear state. Then a constrained energy
minimization is performed with the magnetization direction fixed at
the angle $\theta$ with its original direction.\cite{garkac03prl}
The value of $\theta$ is gradually increased from zero in both directions
for the $3d$ $xy$ model. The corresponding Lagrange multiplier is
an adjustable magnetic field perpendicular to the instantaneous sample's
magnetization. The results of this numerical experiment shown in Fig.
\ref{Fig-dE_and_m_vs_theta_Nalp=2_L=128_HR=1.5}
confirm strong pinning of the F-state. The magnetization does not
rotate easily in the potential landscape created by the RF. Instead,
it exhibits elasticity that can be interpreted as a memory of the
initial state. While the energy increases for the rotation in any
direction, the magnetization decreases to small values for which the
method stops working.

\section{Effects of temperature}

\label{temperature}

\subsection{Numerical method}

\label{sub:MC_method}

At nonzero temperatures we replace the rotation spins towards the
direction of the effective field by Monte Carlo updates,\cite{Gingras-Huse-PRB1996}.
Sequentially at each lattice site, spins are rotated into a random
direction within a cone around the initial spin direction. The width
of the cone increases with the temperature in such a way that approximately
half of new spin directions are accepted. According to the basic Monte
Carlo routine, the new orientation is accepted if the new energy is
lower than the old one. In the opposite case the new orientation is
accepted with the probability $\exp\left[\left(E_{\mathrm{old}}-E_{\mathrm{new}}\right)/T\right]$.
The energy change due to rotation of a single spin is a local quantity
in our model, thus the method is parallelizable and fast.

We combine Monte Carlo spin updates with the energy-conserving over-relaxation
(see Sec. \ref{sec:numerical}), again with probabilities $\alpha$
and $1-\alpha$, respectively. At high temperatures $\alpha=1$ provides
the fastest relaxation to the equilibrium. However, at low temperatures
the method with small $\alpha$ becomes more efficient. This is similar
to the behavior of the relaxation routine used at $T=0$. For large
$\alpha$, the system quickly falls into the nearest shallow energy
minimum (actually a multi-dimensional energy valley) and begins a
long trip along it. For small $\alpha$, the system first finds the
maximum-entropy state that corresponds to a broad and deep energy
basin and then descends into the corresponding minimum.

As the stopping criterion for equilibration, we require that the drift
of the energy value averaged over the interval $n_{\mathrm{avr}}$
of Monte Carlo updates within this interval becomes smaller than the
statistical scatter of the energy. The greater $n_{\mathrm{avr}}$,
the stricter is the equilibration criterion. In the data below we
used $n_{\mathrm{avr}}=30$.

\begin{figure}
\includegraphics[width=8cm]{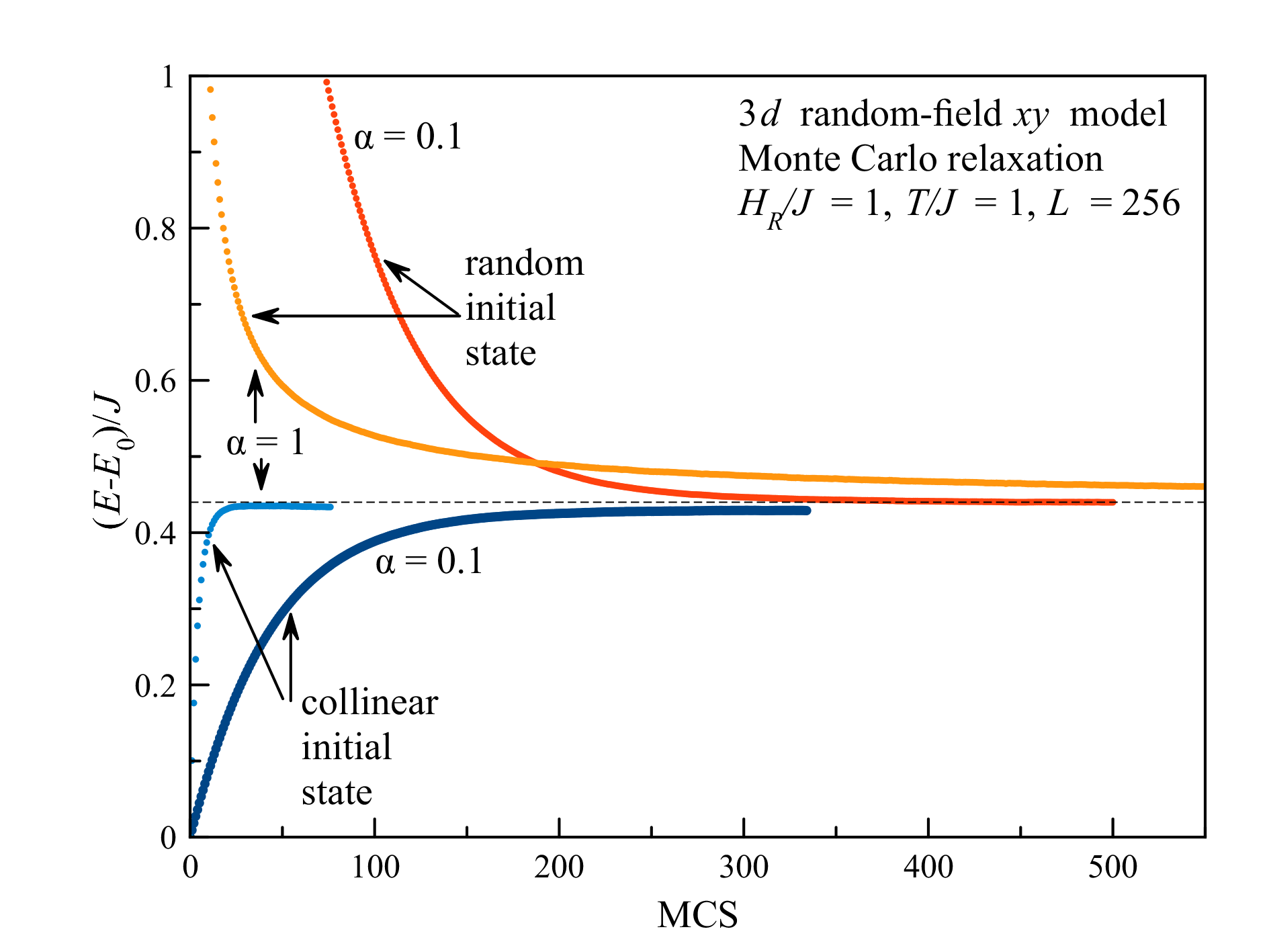}

\caption{Energy relaxation in the Monte Carlo routine.}
\label{Fig-dE_vs_nMC_L=256_HR=1.5_T=1}

\end{figure}

Examples of Monte Carlo energy relaxation out of the collinear and
random initial states for $\alpha=1$ and 0.1 are shown in Fig. \ref{Fig-dE_vs_nMC_L=256_HR=1.5_T=1}.
One can see that for the temperature as low as $T/J=1$ for $\alpha=0.1$
the system relaxes to nearly the same energies from both collinear
and random initial states. On the other hand, for $\alpha=1$ relaxation
out of the random state becomes too slow. At this temperature, states
reached from random and collinear initial conditions are different.
Collinear initial condition results in F states while random initial
condition results in the magnetically disprdered vortex-glass state.
The energy difference between these states is not seen in this scale.
At higher temperatures, the system relaxes into the same magnetically
disordered state out of both random and collinear initial conditions.
This case is non-problematic and the method with $\alpha=1$ is the
fastest.

After the stationary values of the energy and magnetization have been
reached, we typically perform 200 Monte Carlo updates to measure the
physical quantities. Due to the large system size, thermodynamic fluctuations,
as well as the difference between various realizations of the random
field, are small. The statistical scatter seen in the figures is also
small and it is not masking the qualitative features.

\subsection{Numerical results}

\label{sub:MC_results}

\begin{figure}
\includegraphics[width=8cm]{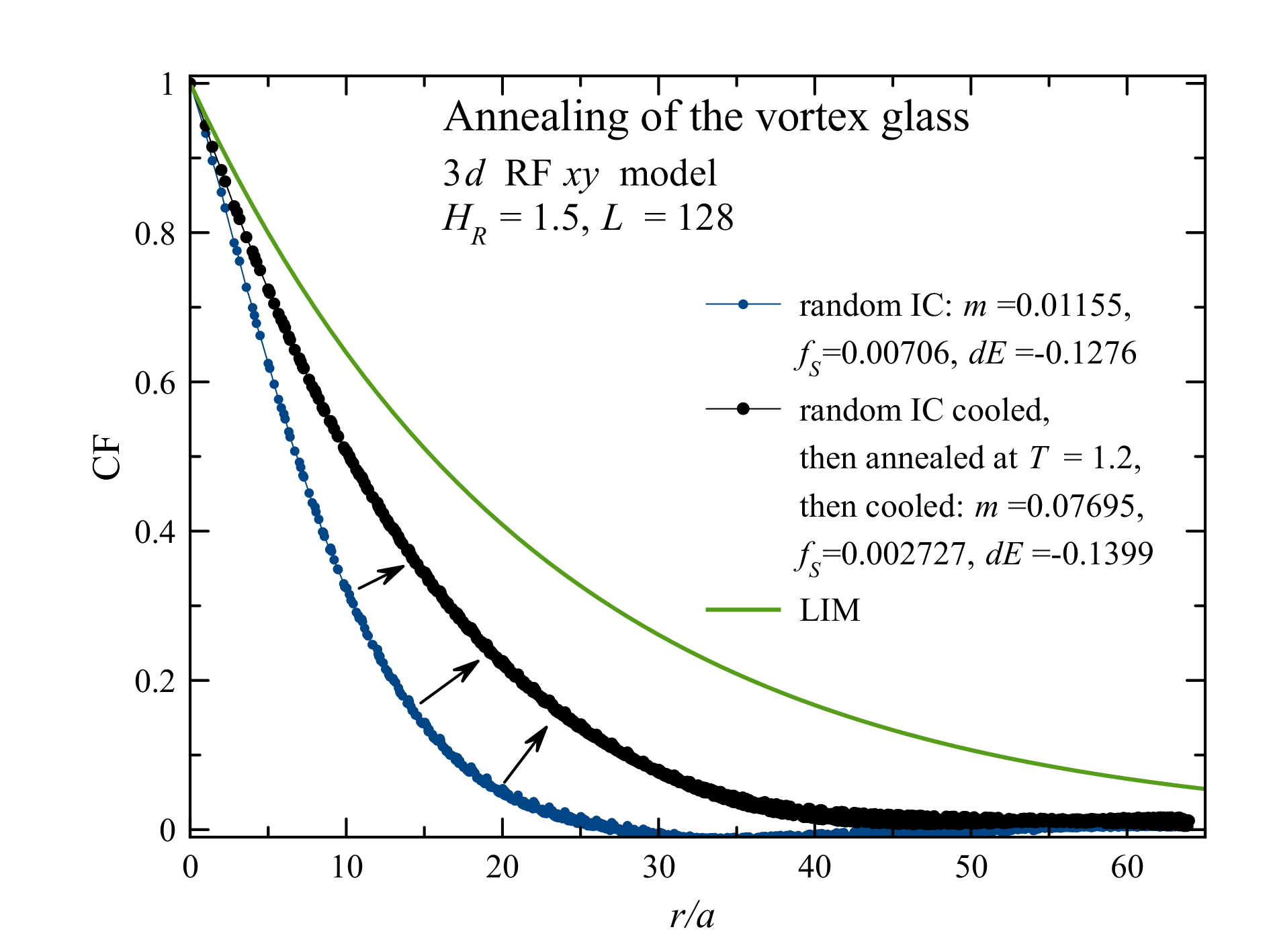}

\caption{Annealing of the vortex-glass state.}

\label{Fig-VG_annealing_CF_vs_n_Nx=Ny=Nz=128_HR=1.5}
\end{figure}

In this Section we consider effects of finite temperature with the
help of the Monte Carlo Metropolis algorithm combined with over-relaxation.
The relevant questions are whether the system orders spontaneously
on cooling and whether finite temperature leads to the full disordering
of the F-state by helping the system to overcome energy barriers.

\begin{figure}
\includegraphics[width=8cm]{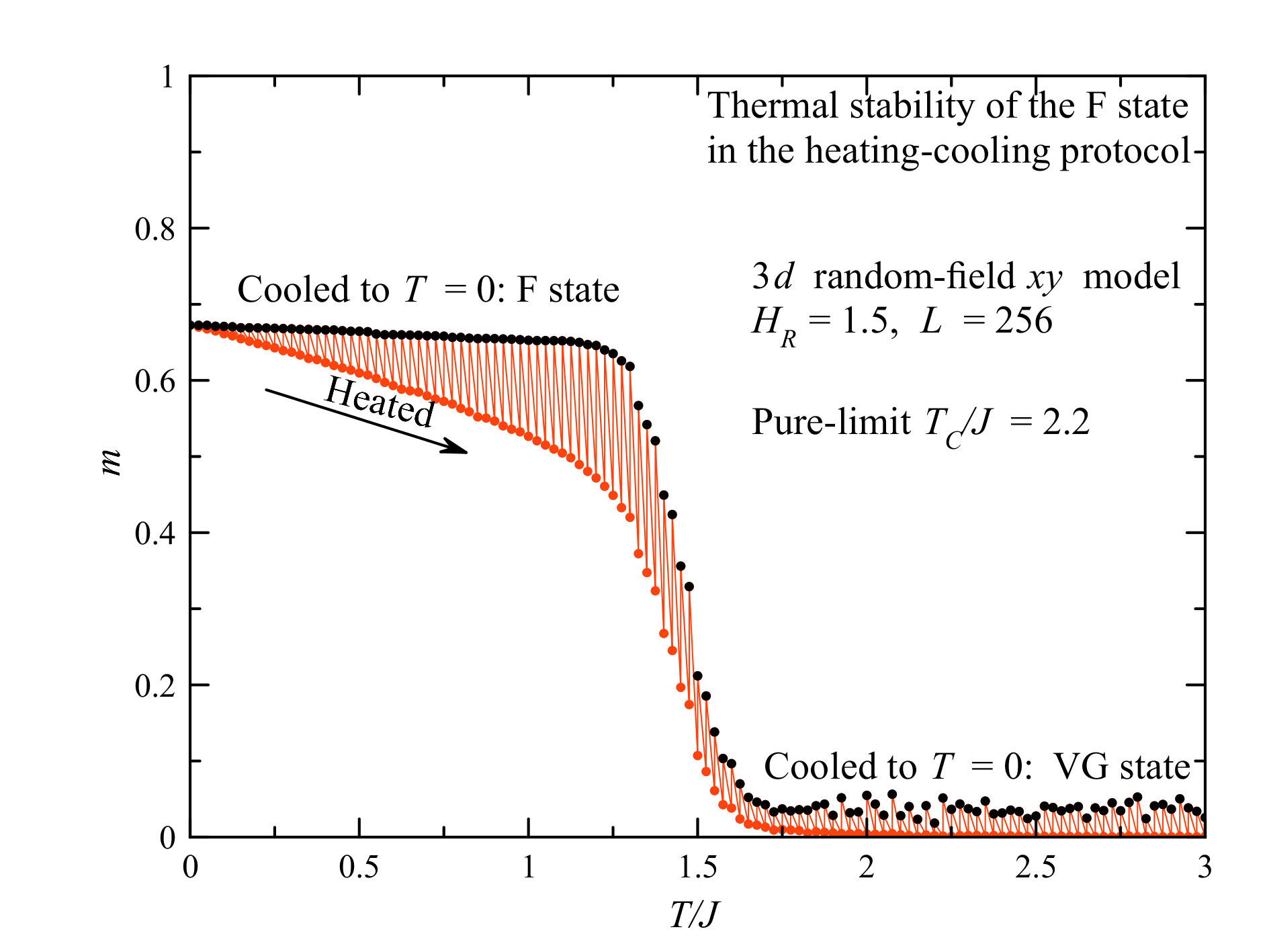}

\caption{Heating-cooling protocol for the $3d$ $xy$ model. The temperature
is increased in steps and each time dropped to zero.}

\label{Fig-m_annealing_vs_T_L=256_HR=1.5_alpha=0.01_alphaMC=0.5_pbc}
\end{figure}

\begin{figure}
\includegraphics[width=8cm]{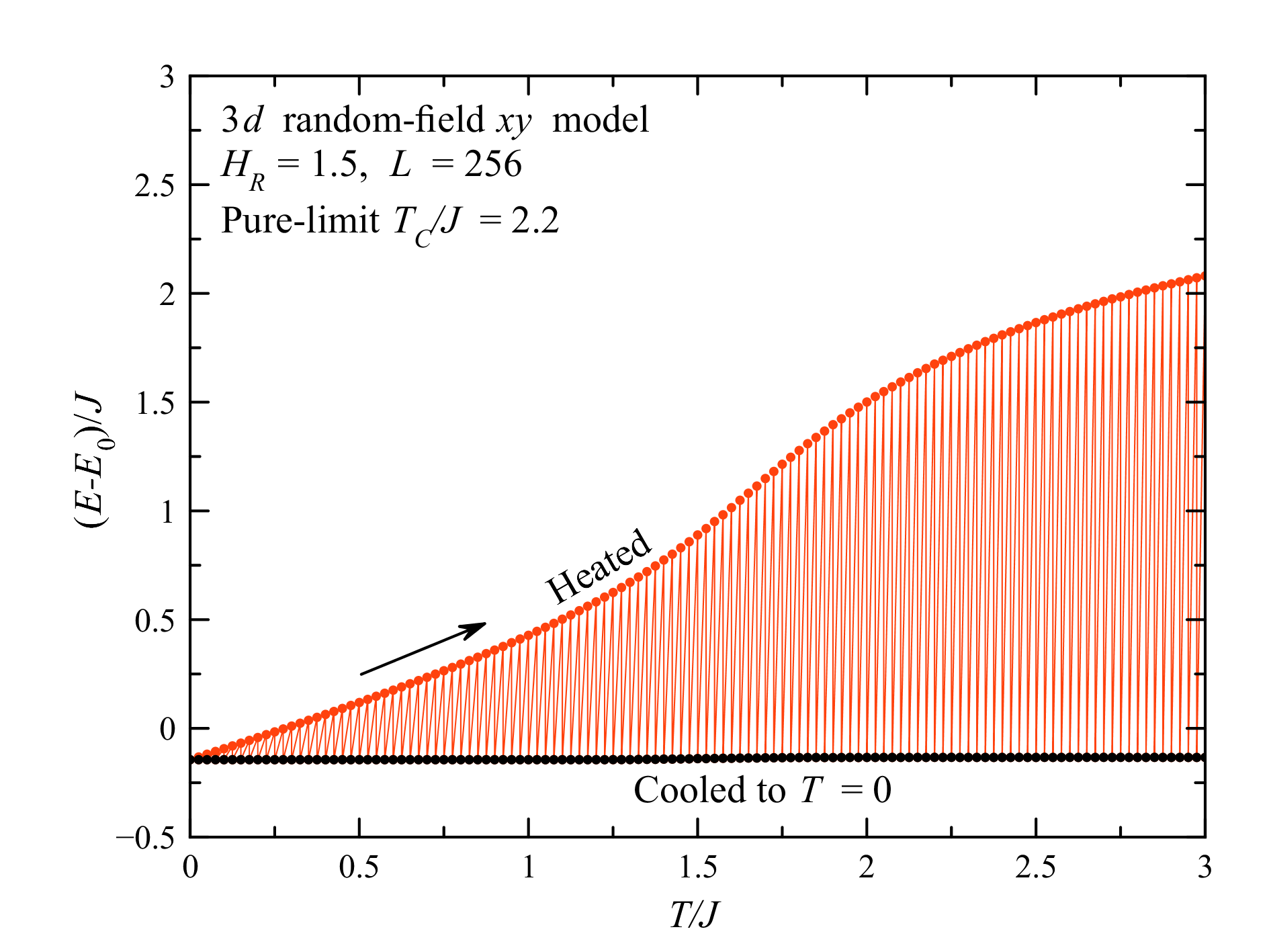}

\includegraphics[width=8cm]{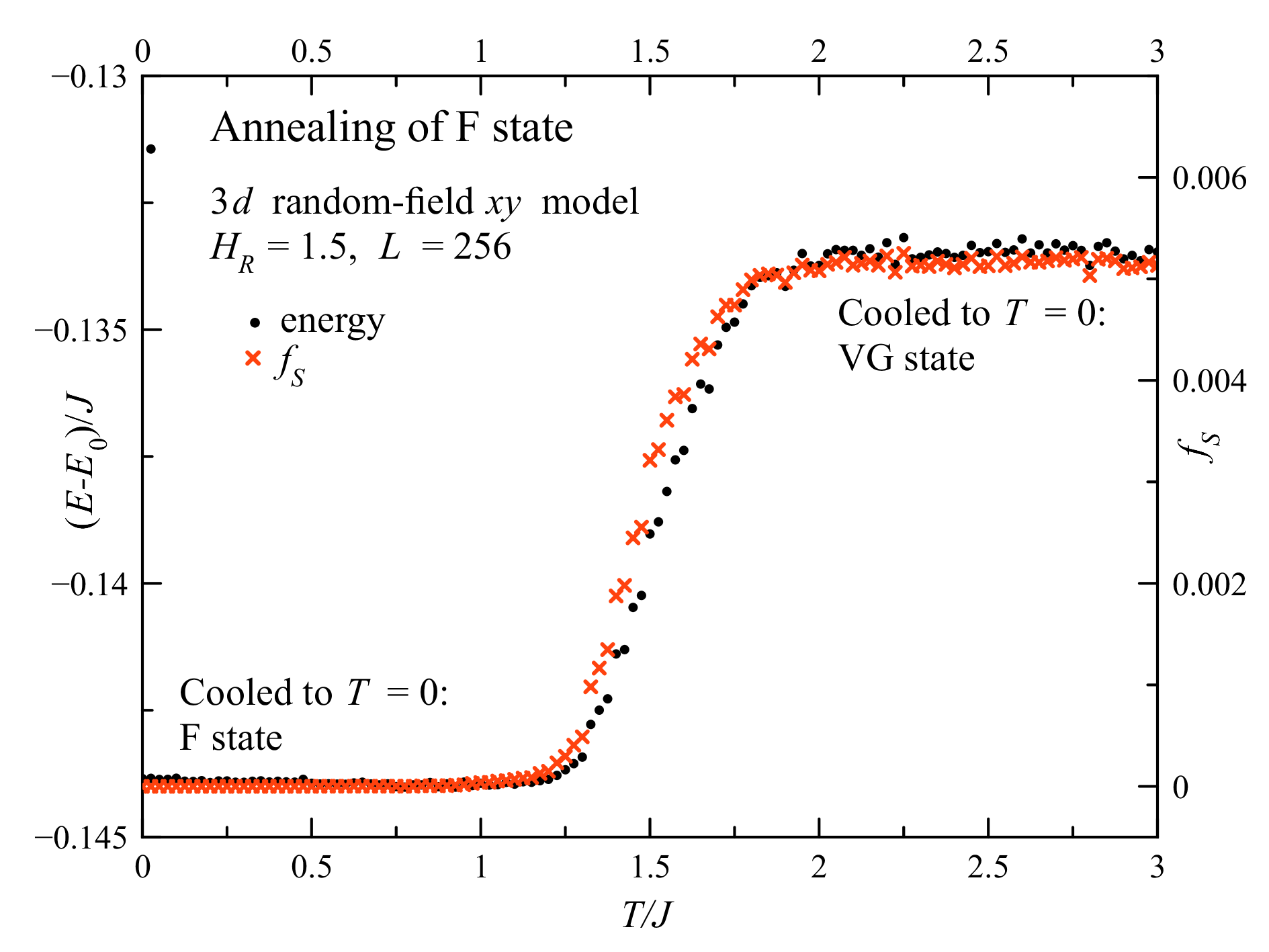}

\caption{Energy in the heating-cooling protocol for the $3d$ $xy$ model.
Lower panel: Magnification of the $T=0$ results.}

\label{Fig-dE_annealing_vs_T_L=256_HR=1.5_alpha=0.01_alphaMC=0.5_pbc}
\end{figure}

The answer to the first question is negative. On cooling, the $3d$
$xy$ system freezes into the vortex-glass (VG) state with a small
magnetization being a finite-size effect. The energy of the VG state
is higher than that of the F-state because of pinned vortex loops.
\cite{GCP-PRB13} Numerical investigation shows that the VG state
is not unique and depends on preparation. Relaxation from a random
state at $T=0$, even using the algorithm with a slow energy loss,
$\alpha\ll1$, leads to the VG states with more vortices and higher
energies. Stepwise lowering temperature leads to the VG states with
less vortices and lower energies. As singularities in the spin field
are breaking spin-spin correlations, the correlation functions of
states with more singularities decay faster. To illustrate this point,
annealing of the VG state of the $3d$ $xy$ model with $H_{R}=1.5$
obtained by relaxation from a random state at $T=0$ has been done.
After initial relaxation at $T=0$, the temperature was raised to
$T/J=1.2$ and the system was equilibrated. After that the temperature
was again dropped to zero. The obtained annealed state has lower vorticity
and lower energy, while its correlation range is longer, as can be
seen in Fig. \ref{Fig-VG_annealing_CF_vs_n_Nx=Ny=Nz=128_HR=1.5}.
Annealing helps to depin and kill some vortex loops, while stronger
pinned loops survive. Heating the system to even higher temperature
would depin more vortex loops. However, this would create new ones.
As a result, the annealing cannot eliminate vortex loops completely
and it cannot bring the system from the VG state to the vortex-free
state.

Next, we investigate thermal stability of the F-state by a heating
protocol that consists of stepwise heating of the initial F-state
to higher temperatures, each time allowing the system to relax at
$T=0$. The temperature sequence has the form $(0,T_{1},0,T_{2},0,T_{3},\ldots)$
with $T_{1}<T_{2}<T_{3}\ldots$. States with non-zero temperature
are obtained with the Metropolis Monte Carlo routine, while zero-temperature
states are obtained by the method of Ref. \cite{GCP-PRB13}
that includes direct rotations of spins toward the effective field
and over-relaxation. Since in this method the fraction of over-relaxation
steps is dominant and the system is relaxing slowly, we do not call
this sequence ``annealing-quenching''. The magnetization results
of this numerical experiment for the $3d$ $xy$ model are shown in
Fig. \ref{Fig-m_annealing_vs_T_L=256_HR=1.5_alpha=0.01_alphaMC=0.5_pbc}.
For the temperatures below 1.2$J$, the system returns to the F-states.
Heating to higher temperatures creates vortex loops that get pinned
and do not collapse upon dropping $T$ to zero. Thus the F-state gets
destroyed by heating to higher temperatures, the resulting state being
the VG state.

The energy in the heating-cooling protocol is shown in Fig. \ref{Fig-dE_annealing_vs_T_L=256_HR=1.5_alpha=0.01_alphaMC=0.5_pbc}.
In the upper panel, the non-zero-temperature branch has the largest
slope (the maximum of the magnetic heat capacity) around $T/J=1.6$
for $H_{R}/J=1.5$ used here. In the absence of the RF there is a
ferromagnetic phase transition at $T/J=2.2$. The $T=0$ branch in
the upper panel looks like a straight line. However, its magnification
in the lower panel shows a fine structure with the crossover from
the F-state to the higher-energy VG state above $T/J=1.2$. The energy
of the VG state perfectly correlates with the fraction of singularities.
\cite{GCP-PRB13}

\begin{figure}
\includegraphics[width=8cm]{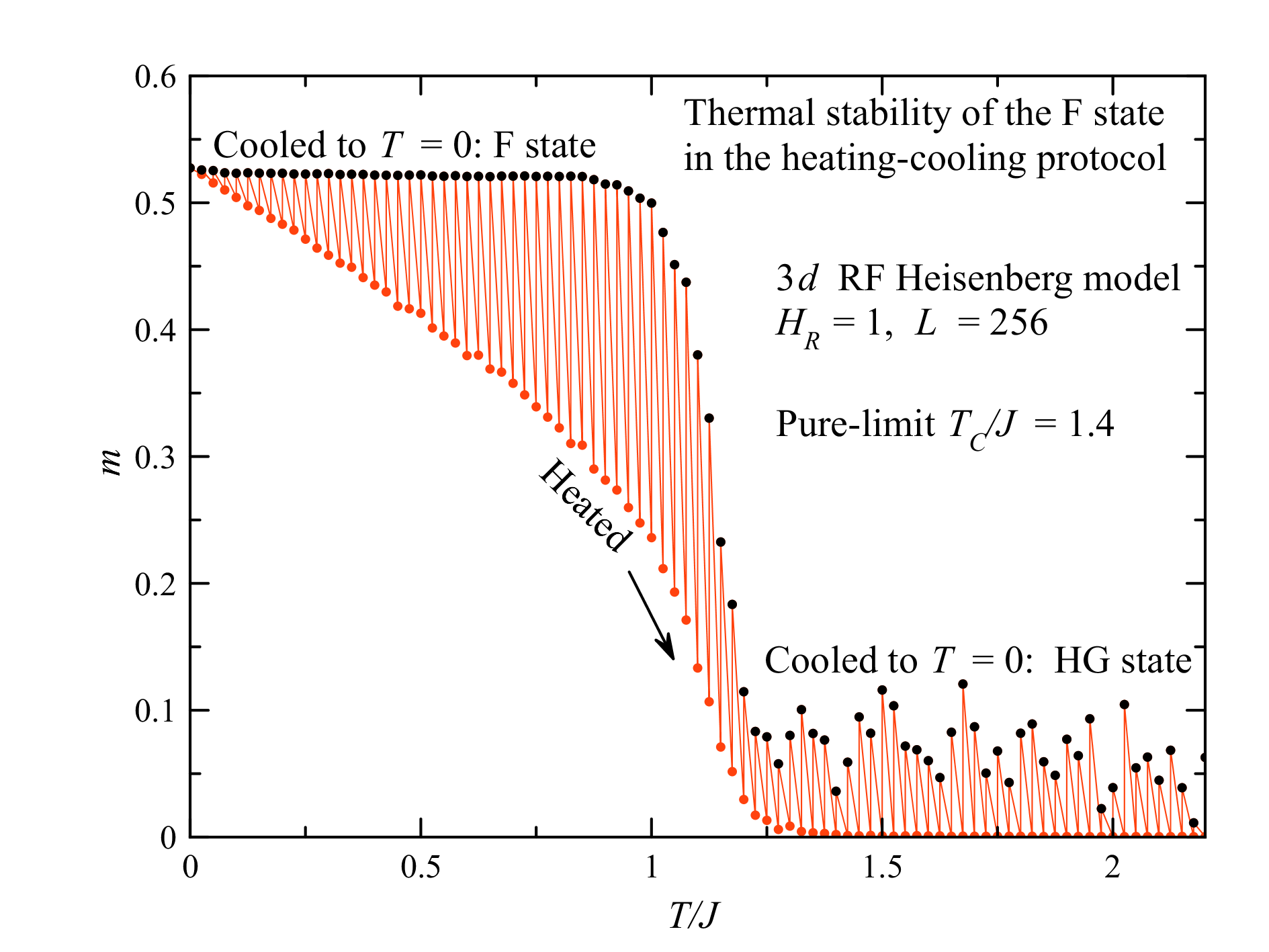}

\includegraphics[width=8cm]{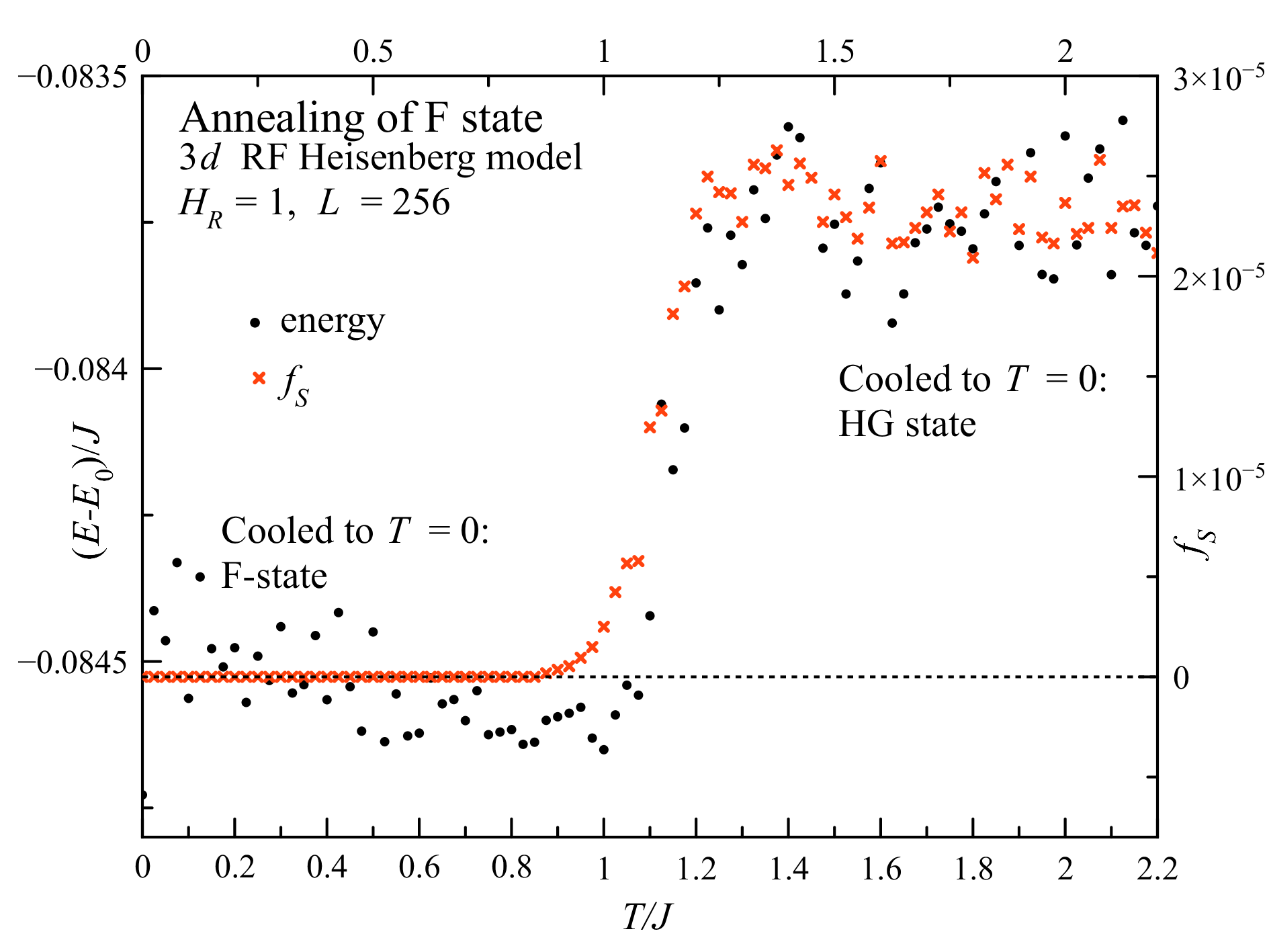}

\caption{Heating-cooling protocol for the $3d$ Heisenberg model. Lower panel:
Energy and singularity fraction at $T=0$.}

\label{Fig-m_annealing_vs_T_3d_H_L=256_HR=1_pbc}
\end{figure}

We also have performed the heating-cooling experiment for a $3d$
Heisenberg model. The results shown in Fig. \ref{Fig-m_annealing_vs_T_3d_H_L=256_HR=1_pbc}
are similar to those for the $xy$ model above. Since hedgehogs in
the Heisenberg model carry much less energy than the vortex loops
in the $xy$ model, the energy of the hedgehog-glass (HG) is only
slightly above that of the F-state.

\section{Discussion and Conclusions}

\label{conclusions}

We have presented a comparative study of glassy states of three-dimensional
$xy$ and Heisenberg random-field models. The $xy$ spin model is
conceptually similar to the model of a pinned flux lattice, even though
the symmetry of the two models is slightly different. \cite{Gingras-Huse-PRB1996}
The Heisenberg RF model has practical implementation in antiferromagnets
with quenched disorder. \cite{Fishman} Some of the properties of
the two models may be also relevant to properties of amorphous and
sintered magnets, although there are very significant differences
between the effects of the random field and random anisotropy.

Earlier we have found that the properties of the random-field model
are controlled by topology. \cite{PGC-PRL2014} For the $n$-component
spin in $d$ dimensions the reversible behavior with exponential decay
of correlations occurs at $n>d+1$ when topological defects are absent.
At $n\leq d$ the random-field system possesses pinned topological
defects and exhibits glassy behavior. The focus of this paper has
been on the properties of the F-state obtained by relaxation from
the initially ordered state. One of our main findings is a profound
difference between the F-states of the $xy$ model and the Heisenberg
model. While the latter can be transformed into a lower energy state
with $m=0$ by applying an appropriate hysteresis cycle, the $xy$
F-state appears robust against attempts to lower its energy and decrease
the magnetization at the same time. This can be traced to the inevitable
presence of topological defects in the $m=0$ state at $n\leq d$.
\cite{GCP-PRB13,PGC-PRL2014} In the $3d$ Heisenberg model such defects
are hedgehogs that have relatively low energy as compared to the vortex
loops in the $3d$ $xy$ model. It is the reluctance of the $xy$
RF system to form high-energy vortex loops that makes the F-state
robust with respect to disordering.

Our other interesting finding is that the F-state possesses memory
of the initial state. There is an infinite number of F-states that
differ from each other by the direction of the global magnetic moment
determined by the initial ordered state that the F-state has evolved
from. Unlike in a ferromagnet with the exchange interaction only,
these states are separated by energy barriers due to the random field.
Small rotation of the magnetization of the F-state reveals its elasticity:
rotation generates forces that return ${\bf m}$ to its original direction.
This makes one wonder whether the F-state has any relevance to the
elastic glass discussed in the past. \cite{Nattermann-2000,Fisher-PRL1997}

We have also studied the effect of temperature on the F-state. Naively,
one would expect that it can be easily annealed towards thermal equilibrium.
Our results do not support this expectation. While heating helps to
depin topological defects, it also creates the new ones. Annealing
of the vortex glass state does result in the lower vorticity and lower
energy but it cannot eliminate vortices completely. On the contrary,
heating and cooling of the F-state generates vortex loops that have
been completely absent at the beginning. This reduces the magnetization
but also leads to the higher energy than the energy of the F-state
not subjected to the heating-cooling procedure. The bottom line is
that the RF system, when cooled down after exposure to elevated temperature,
behaves as a glass. It stores the thermal energy in the form of pinned
topological defects.

There is a deep analogy between this state and the state of a conventional
bulk ferromagnet with defects. In the case of the conventional ferromagnet
the magnetization and the internal dipolar field adjust self-consistently
through equations of magnetostatics. In a similar manner, the magnetization
of the RF system adjusts self-consistently to the average random field.
When cooled in a zero field below the Curie temperature, a conventional
ferromagnet ends up in a state with small domains and high energy
dominated by pinned domain walls. Similarly, the RF magnet ends up
with many pinned topological defects: vortex loops in the $xy$ model
and hedgehogs in the Heisenberg model. When relaxing from a magnetized
state at a sufficiently low temperature, a conventional ferromagnet
ends up in a state with a finite magnetization because domain walls
cannot easily penetrate in the magnet due to pinning and, thus, cannot
form the $m=0$ ground state. In a similar manner, the RF magnet ends
up in the F-state when it relaxes from the ordered initial state.
There are significant differences though as well. The $m=0$ ground
state of a conventional ferromagnet is the minimum of magnetostatic
energy that has not been considered in the context of the RF system
studied by us and by other authors. It is not at all obvious whether
the $xy$ system in the F-state can lower its energy by accommodating
a large vortex loop that would reduce the magnetization to zero. Although
there may be little profit in discussing the ground state of the glassy
system, some remarks are in order.

The Imry-Ma argument is clearly valid at the qualitative level as
long as the $m=0$ state does not require topological defects. \cite{PGC-PRL2014}
At $n\leq d$ the argument becomes less obvious as the IM domains
cannot be formed without topological defects. On the first glance,
it appears that topological defects only modify the IM argument but
do not destroy it. Indeed, in a $3d$ $xy$ model there would be roughly
one vortex loop per IM domain of size $R$. The energy of the loop
would be of order $2\pi Js^{2}(R/a)\ln(R/a)$, that corresponds to
the energy $2\pi Js^{2}(a/R)^{2}\ln(R/a)$ per spin. This modifies
the exchange energy in Eq. (\ref{eq:Imry-Ma}) by a factor proportional
to $\log(R)$ and leads to a greater but still finite $R_{f}$ in
the disordered state. The energy lowering due to the formation of
the IM state becomes strongly reduced when vortex loops are present,
see Eq. (\ref{eq:EIM_vortex_loops}). At the same time, the lowering
of the energy in the vortex-free F-state, see Eq. (\ref{eq:Imry-Ma-phi})
and below, remains of the same order as the regular IM lowering decribed
by Eq. (\ref{eq:E-IM-d}).

For the $3d$ Heisenberg model, the energy of a hedgehog in the domain
of size $R$ is $4\pi Js^{2}(R/a)$, which modifies the exhange energy
per spin and the resulting IM energy by a factor of order unity. This
makes estimates of the energies of the disordered IM state with hedgehogs
and the F-states without hedgehogs the same. Making hysteresis loops
with decreasing amplitude converts the F-state into a disordered state
of the lower energy. This indicates that the ground state of the $3d$
Heisenberg model may be disordered in spite of hedgehogs.

One should note that in the presence of topological defects the IM
argument becomes less precise as it ignores misalignment of the spin
field with the average random field due to defects, as well as the
interaction between topological defects. There is also a scaling argument
that makes the IM argument even less obvious. Indeed, the IM argument
relies on the large size of IM domains in which the direction of the
magnetization follows the RF field averaged over the volume of the
domain. It implies smooth rotation of the magnetization from one domain
to the other. The argument would not apply to the case in which the
RF field $h$ at each cite of a $3d$ cubic lattice is of the order
of the exchange field $6Js$ created by the neighboring spins, because
it would be difficult to say without direct computation whether the
effect of the RF would win over the effect of the exchange. In this
connection, one should notice that by considering blocks of spins
of size $r$ satisfying $a<r<R_{f}$, the original problem described
by the Hamiltonian (\ref{eq:ham-discrete}) can be rescaled to the
problem described by the same Hamiltonian with the rescaled $s_{r}=s(r/a)^{3}$,
$J_{r}=J(a/r)^{5}$, and $h_{r}=h(a/r)^{3/2}$. This gives the same
expression for the correlation length, $R_{f}/r=(2\pi/9)(1-1/n)^{-1}(6J_{r}s_{r}/h_{r})^{2}$.
For the blocks of size $r\sim R_{f}$ one has $h_{r}\sim6J_{r}s_{r}$
in the rescaled problem. This shows that the existence of a small
parameter $h/(6J)$ in the original problem with weak RF may be an
illusion. The original problem with $h\ll6Js$ is mathematically equivalent
to the rescaled problem in which the local exchange field and the
RF field are of the same order of magnitude. At $h\gg6Js$ the spins
obviously align with the RF, yielding the state with $m=0$. The question,
therefore, is whether decreasing $h$ from $h\gg6J$ would result
in the bifurcation or the ground state to a non-zero $m$ at some
$h\sim6Js$. This is suggested by Figs. \ref{Fig-m_vs_HR_L=512}
and \ref{Fig-m_vs_HR_n=3_L=512_pbc}.
The answer for the ground state may depend on the number of spin components
$n$. However, looking for the ground state in a glassy system requires
a different numerical algorithm based on the global energy minimization.
Whether this is worth the effort is another question.

\section{Acknowledgements}

This work has been supported by the Grant No. DE-FG02-93ER45487 funded by U.S. Department of Energy, Office of Science.

\end{document}